\documentclass[journal]{IEEEtran}
\usepackage{mathtools}
\usepackage{amsmath} 
\usepackage{amssymb}
\usepackage{amsfonts}
\usepackage{verbatim} 
\usepackage{bm,color,soul}
\usepackage{algorithm,algorithmic}
\usepackage{cite}
\usepackage{caption}
\usepackage{subfigure}
\usepackage{stfloats} 
\usepackage[colorlinks, linkcolor=red, anchorcolor=black, citecolor=green]{hyperref} 
\usepackage{graphicx} 
\usepackage{subeqnarray}
\usepackage{cases}
\usepackage{makecell} 
\usepackage{enumerate}
\usepackage{array}
\usepackage{url}
\usepackage{balance}

\newtheorem{rem}{Remark}
\newtheorem{prop}{Proposition}
\newtheorem{lem}{Lemma}

\newcommand{\rnum}[1]{\uppercase\expandafter{\romannumeral #1\relax}}
\ifCLASSINFOpdf
\else
\fi
\makeatletter
\renewcommand{\maketag@@@}[1]{\hbox{\m@th\normalsize\normalfont#1}}
\makeatother

\IEEEoverridecommandlockouts

\begin{document}
%
\title{Active IRS-Enabled Integrated Sensing and Communications with Extended Targets\vspace{-0.0em} }

\author{\IEEEauthorblockN{Yuan Fang,~\IEEEmembership{Member,~IEEE}, Xianxin Song,~\IEEEmembership{Member,~IEEE}, Huazhou Hou,~\IEEEmembership{Member,~IEEE}, Ziguo Zhong,\\ Xianghao Yu,~\IEEEmembership{Senior Member,~IEEE}, Jie Xu,~\IEEEmembership{Fellow,~IEEE}, and Yongming Huang,~\IEEEmembership{Fellow,~IEEE}\vspace{-8mm} 
	}
	\thanks{Y. Fang is with the Purple Mountain Laboratories, Nanjing  211111, China, and the Department of Electrical Engineering, City University of Hong Kong, Hong Kong, China (e-mail: fangyuan94@outlook.com).}
	\thanks{X. Song and X. Yu are with the Department of Electrical Engineering, City University of Hong Kong, Hong Kong, China (e-mail: xianxin.song@cityu.edu.hk; alex.yu@cityu.edu.hk).}
	\thanks{H. Hou and Z. Zhong are with Purple Mountain Laboratories, Nanjing  211111, China (e-mail: houhuazhou@pmlabs.com.cn; zhongziguo@pmlabs.com.cn).}
	\thanks{J. Xu is with School of Science and Engineering, Shenzhen Future Network of Intelligence Institute (FNii-Shenzhen), and Guangdong Provincial Key Laboratory of Future Networks of Intelligence, The Chinese University of Hong Kong, Shenzhen, Guangdong 518172, China (e-mail: xujie@cuhk.edu.cn).} 
	\thanks{Y. Huang is with National Mobile Communications Research Laboratory, School of Information Science and Engineering, Southeast University, Nanjing 210096, China, and also with the Purple Mountain Laboratories, Nanjing 211111, China (e-mail: huangym@seu.edu.cn).}
	\thanks{X. Yu is the corresponding author.}
}

\maketitle
\begin{abstract}
This paper studies the active intelligent reflecting surface (IRS)-enabled integrated sensing and communications (ISAC), in which an active IRS is deployed to assist the base station (BS) in serving multiple communication users (CUs) and simultaneously sensing an \emph{extended} target at the non-line-of-sight (NLoS) area of the BS. The active IRS has the capability of amplifying the reflected signals so as to overcome significant reflection path loss in NLoS communication and sensing. In particular, we derive the sensing  Cram\'{e}r-Rao bound (CRB) for estimating the target response matrix. Accordingly, we jointly optimize the transmit beamforming at the BS and the reflective beamforming at the active IRS to minimize the sensing CRB, subject to the signal-to-interference-plus-noise ratio (SINR) requirements at the CUs, the  transmit power budgets at the BS and active IRS, as well as the power amplification gain constraints at the active IRS. The CRB minimization problem is highly non-convex and thus difficult to solve in general. To address this challenge, we first focus on two specified conditions by considering the sensing-only scenario via ignoring the SINR constraints for communications, for which the closed-form optimal transmit beamforming is derived. Then, we propose two efficient alternating optimization (AO)-based algorithms to obtain high-quality solutions for the general ISAC scenarios. Next, we analyze the inherent relationship between the power scaling at the BS and the amplification scaling at the active IRS. It is shown that the active IRS always amplifies the signal using the maximum amplification gain under practical system settings. Finally, numerical results are provided to verify the effectiveness of the proposed AO-based algorithms and the benefits of active IRS-enabled ISAC compared to its passive IRSs  counterparts.
\end{abstract}


%
\IEEEpeerreviewmaketitle

\vspace{-3mm}
\section{Introduction}
Integrated sensing and communications (ISAC) has emerged as a pivotal technology for sixth-generation (6G) wireless networks, where the integration of sensing and communication functionalities within shared infrastructure and spectrum resources is anticipated to revolutionize wireless services \cite{cui2021integrating,zhang2021overview}. By leveraging the wireless sensing capability, cellular base stations (BSs) can extract valuable environmental and object information from echo signals \cite{liu2022integrated}. However, the effectiveness of ISAC networks intrinsically relies on the wireless propagation environment. For instance, the accuracy of wireless sensing heavily relies on the availability of line-of-sight (LoS) paths, which are essential for precise target angle and range estimation. Simultaneously, wireless communication performance relies on the strength of the communication channels and the interference level, which are critical for meeting stringent quality-of-service (QoS) and signal-to-interference-plus-noise ratio (SINR) requirements of communication users (CUs). However, in practical scenarios, the presence of obstacles such as buildings, trees, furniture, and walls can significantly attenuate or completely block the propagation links between transceivers \cite{huang2019reconfigurable}. This environmental complexity poses substantial challenges to the realization of high-performance ISAC, necessitating advanced solutions to mitigate signal degradation and ensure reliable service delivery \cite{hua2022joint,hua2023intelligent}.

Intelligent reflecting surfaces (IRSs) are a kind of meta-surfaces composed of numerous  electromagnetic elements, each capable of dynamically adjusting the phase shifts of incident signals. By intelligently adjusting these reflections, IRSs can establish favorable virtual LoS links between transmitters and receivers, thereby bypassing the blockage \cite{di2020smart,wu2021intelligent}. Hence, the application of IRSs has shown remarkable potential in improving coverage and increasing signal quality, thus enhancing the performances of wireless communication \cite{yu2021irs,wu2023intelligent,xu2020resource}, sensing \cite{song2024cramer}, and ISAC \cite{Xianxin2025overview}.

Deploying IRSs into ISAC systems not only improves communication performance but also significantly enhances the sensing capability, especially when direct LoS paths between BS and targets/CUs are obstructed \cite{song2024cramer}. In particular, the authors of \cite{jiang2021intelligent} and \cite{yan2022reconfigurable} incorporated IRSs into ISAC systems to provide additional propagation paths for radar echo signals, while simultaneously enhancing communication performance. With implementing an IRS into ISAC systems, the work \cite{liu2022joint} jointly designed the transmit waveform, receive filter, and IRS reflection coefficients in cluttered environments. In \cite{wang2021joint1} and \cite{wang2021joint2}, the IRS was deployed in ISAC systems to mitigate inter-user interference while ensuring the radar sensing beam pattern and Cram\'{e}r-Rao bound (CRB) constraints. In \cite{song2023intelligent}, an IRS was adopted to create virtual LoS links, enabling the estimation of potential targets blocked by obstacles. The work \cite{chen2023ris} explored a bi-static sensing system assisted by passive IRSs, where the angle information of perceived objects was estimated using the BS-target-IRS-sensor link.  In \cite{zhang2022joint}, the authors examined the trade-off between maximizing communication data rate and sensing mutual information. Additionally, unlike previous studies focused on narrowband scenarios, recent works \cite{wei2022multiple} and \cite{wei2022irs} explored IRS-aided wideband ISAC systems with orthogonal frequency division multiplexing (OFDM) technology. Taking into account the challenges posed by imperfect angle knowledge and channel state information (CSI), \cite{luan2023robust}  studied robust beamforming design for IRS-aided ISAC systems. Furthermore, the multi-user sum-rate maximization under radar signal-to-noise ratio (SNR) or CRB constraints was investigated in \cite{liu2024ris}.  However, all of the aforementioned works focused on passive IRSs, where system performance is constrained by the multiplicative fading effect inherent to IRSs \cite{buzzi2022foundations}. This effect leads to substantial signal attenuation, especially when the receiver is not in close proximity to the IRS, resulting in marginal performance improvements.

To overcome the limitations of passive IRSs, {\it active} IRS technology has been proposed \cite{zhang2022active}. Active IRSs integrate reflection-type amplifiers into existing passive electromagnetic components, enabling not only reflection but also amplification of incident signals \cite{kang2024active}. This amplification capability effectively mitigates the multiplicative fading issue, thus significantly enhancing the ISAC performance. Recently, in \cite{salem2022active}, an active IRS was utilized to enhance the communication secrecy rate while ensuring a minimum radar detection SNR. Furthermore, \cite{zhang2022CRAN} explored active IRSs in an ISAC cloud radio access network, where an active IRS was deployed to address the blockage issue between BS and targets/users. The radar beam pattern towards the sensing targets was optimized to boost the sensing performance. Similarly, the work in \cite{zhu2023joint} also adopted an active IRS to create an additional virtual LoS link between the BS and the target, where the transmit/receive and reflective beamforming were jointly optimized to maximize the radar SNR while meeting the SINR requirements for CUs. In addition,  \cite{yu2023active} focused on maximizing radar SINR in an active IRS-aided ISAC system while ensuring the communication SINR constraints of CUs. Furthermore, in \cite{zhu2024cramer}, the authors investigated active IRS-assisted ISAC systems by minimizing the CRB for point target estimation through a joint design of transmit and reflective beamforming. 

Nevertheless, previous works on active IRS-enabled ISAC mainly focused on point targets, which are typically modeled as single and isolated points. In contrast, extended targets, such as vehicles or groups of objects, are usually modeled as entities with significant size and spatial extent \cite{koch2008bayesian}. The study of active IRS-enabled ISAC systems in the context of extended targets still remains unexplored. Motivated by this, we investigate the active IRS-enabled ISAC system with extended targets, where an active IRS is employed to assist the BS in serving multiple CUs while sensing an extended target located in the non-LoS (NLoS) area of the BS. The active IRS can amplify reflected signals, thereby mitigating the severe reflection path loss typically encountered in NLoS communication and sensing.  The main contributions and results of this paper are summarized as follows.
\begin{itemize}
	\item First, we derive the closed-form CRB for estimating the target response matrix of an extended target. Based on the derived CRB, we aim to minimize the estimation CRB to improve sensing performance. In particular, we jointly optimize the transmit beamforming at the BS and the reflective beamforming at the active IRS to minimize the CRB, subject to the SINR requirements at the CUs, the constraint on the maximum transmit power at the BS, as well as the maximum transmit power and amplification gain constraints at the active IRS. 
	\item Then, we consider the sensing-only scenario by ignoring the SINR constraints at CUs. In this case, we first derive {\it closed-form} optimal transmit beamforming at the BS under two specified conditions. By exploiting the structure of the closed-form CRB, we propose an alternating optimization (AO) algorithm to obtain the joint beamforming design. In particular, we obtain the optimal transmit beamforming design with a given reflective beamforming design, and optimize the phase shifts and amplification under a given transmit beamforming deign alternately by using semi-definite relaxation (SDR) and successive convex approximation (SCA) techniques, respectively. 
    \item \textcolor{black}{Next, we consider the general ISCA scenario and propose an AO algorithm to obtain the joint beamforming design. In particular, we first derive the optimal beamforming for both information and sensing signals using SDR with a given reflective beamforming design. Subsequently, we alternately optimize the phase shifts and amplification at the active IRS to maximize the minimum SINR among all CUs and minimize the estimation CRB, respectively.} 
	\item Subsequently, we analyze the scaling law of the transmit beamforming at the BS and the amplification coefficients at the active IRS. We reveal the relationship between the optimal scaling of the transmit beamforming and the amplification coefficients, and find out that the active IRS always amplifies the signal using the maximum amplification gain under practical system settings.
	\item Finally, numerical results verify the effectiveness of our proposed AO-based algorithms and the advantages of active IRS-enabled ISAC compared to that with passive IRSs. It is shown that the proposed \textcolor{black}{AO-based designs} outperform various benchmark schemes with transmit beamforming only, reflective beamforming only, and zero-forcing (ZF) beamforming. 
\end{itemize}


	{\it Notations:} The circularly symmetric complex Gaussian distribution with mean $\bm{\mu}$ and covariance $\mathbf{A}$ are denoted as $\mathcal{CN}(\bm{\mu},\mathbf{A})$. The notations $(\cdot)^{T}$, $(\cdot)^{*}$, $(\cdot)^{H}$, and $\mathrm{tr}(\cdot)$ denote the transpose, conjugate, conjugate-transpose, and trace operators, respectively. $\mathbf{I}_{L}$ stands for the identity matrix of size $L \times L$. $\Re(\cdot)$ and $\Im(\cdot)$ denote the real and imaginary parts of the argument, respectively. $|\cdot|$ and $\mathrm{arg}\left\{\cdot\right\}$ denote the absolute value and angle of a complex element, respectively. $\mathrm{vec}(\cdot)$ denotes the vectorization operator, $\mathbb{ E}(\cdot)$ denotes the expectation operation, $\mathrm{diag}(\mathbf{x})$ forms a diagonal matrix with the diagonal entries specified by vector $\mathbf{x}$, $\mathrm{Diag}(\mathbf{X})$ represents a diagonal matrix with the diagonal entries specified by the diagonal elements in $\mathbf{X}$, and $\mathrm{Diagvec}(\mathbf{X})$ denotes a vector with its entries specified by the diagonal elements in $\mathbf{X}$. $\mathbf{x}^a$ represents the operation of each element in the vector to the power of $a$. $\mathrm {rank}\left(\mathbf{X}\right)$ denotes the rank of matrix $\mathbf{X}$ and $[\cdot]_{i,j}$ denotes the $(i,j)$-th element of a matrix. $\jmath$ denotes the imaginary unit. $\otimes$ and $\circ$ denote the Kronecker product and Hadamard product operators, respectively. The notations $\max \{ \cdot \}$ and $\min \{ \cdot \}$ are the operation of obtaining the maximum and minimum values from an array, respectively.

\section{System Model}
 \begin{figure}[htbp]
 	\setlength{\abovecaptionskip}{-0pt}
 	\setlength{\belowcaptionskip}{-5mm}
 	\centering
 	\includegraphics[width=0.4\textwidth]{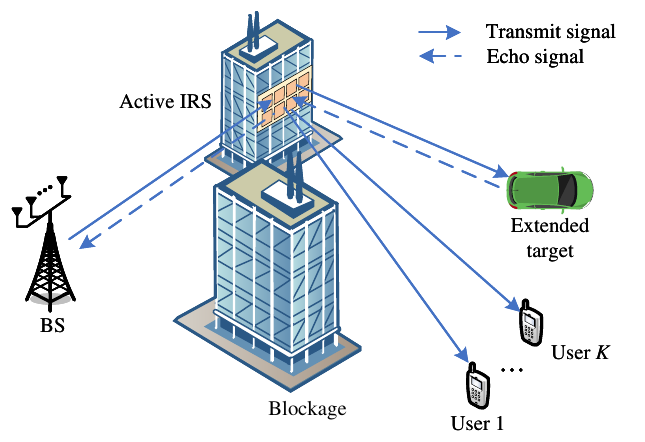}
 	\caption{The active IRS-enabled ISAC.}\label{fig:SystemModel}
 \end{figure}
Consider an active IRS-enabled ISAC system as shown in Fig. \ref{fig:SystemModel}, where an active IRS is deployed to assist the BS in serving  $K$ single-antenna CUs that would otherwise be blocked while simultaneously sensing an extended target. The BS and IRS are equipped with $M$ uniform linear array (ULA) antennas and $N$ ULA reflecting elements, respectively. Let $\mathcal{M} = \{1,\ldots,M\}$ denote the set of antennas at the BS, $\mathcal{N} = \{1,\ldots,N\}$ denote the set of reflecting elements at the active IRS, and $\mathcal{K}=\{1,\ldots,K\}$ denote the set of CUs in this system, respectively. The ISAC dwell time set is denoted by $\mathcal{T}=\{1,\ldots,T\}$.

We consider a quasi-static channel model, where the wireless channels are assumed to remain constant during the ISAC dwell time $\mathcal{T}$. The matrix $\mathbf{G} \in \mathbb{C}^{N \times M}$ represents the channel between the BS and IRS, and vector $\mathbf{h}_{k} \in \mathbb{C}^{N\times 1}$ denotes the channel vector between the IRS and CU $k, \forall k \in \mathcal{K}$. These channels can be accurately estimated using established channel estimation techniques, such as those discussed in \cite{zheng2022survey}. The active IRS is assumed to be carefully positioned to maintain LoS conditions between itself and the target. In this case, the IRS reflects signals transmitted from the BS or echo signals from the target, with the BS responsible for receiving and processing the sensing signals.

Next, we present the design of transmit beamforming at the BS and the reflective beamforming at the IRS. Let $\bm{\psi} = [\psi_{1},\ldots, \psi_{N}]^{T}$ denote the complex reflection coefficients imposed by the IRS, and $a_{\text{max}}$ denote the maximum magnitude amplification gain of the elements at the IRS, respectively. Since each element of the active IRS can tune both the phase and magnitude of the signal, the complex reflection coefficient can be expressed as $\psi_{n} = a_{n}e^{\jmath\rho_{n}}, \forall n \in \mathcal{N}$, where $a_{n}$ and $\rho_{n}$ denote the magnitude  and phase of the reflection coefficient, respectively \cite{zhang2022active}. The maximum power amplification gain constraints at the active IRS are given by $|\psi_{n}|=a_{n} \leq a_{\text{max}}, \forall n\in \mathcal{N}$. At each time symbol $t\in\mathcal{T}$, let ${s}_{k}[t]$ denote the transmit information signal for CU $k$ with ${s}_{k}[t] \sim \mathcal{CN}\left(0,1\right)$, and $\mathbf{w}_{k}$ denote the corresponding information beamforming vector. Let $\mathbf{s}_{0}[t] \sim \mathcal{CN}\left(\mathbf{0},\mathbf{R}_{0}\right)$ denote the dedicated sensing signal, which is randomly generated and independent of ${s}_{k}[t]$ with $\mathbf{R}_{0}$ being the convariance matrix of sensing signal.  Then, the transmit signal from the BS at time symbol $t$ is given by $\mathbf{x}[t] = \sum\nolimits_{k\in \mathcal{K}}\mathbf{w}_{k}s_{k}[t]+\mathbf{s}_{0}[t]$,
and the sample covariance matrix of the transmit signal from the BS over the dwell time symbols is given by
\begin{align}
	\mathbf{R}_{\mathbf{x}} &=\! \frac{1}{T} \sum\nolimits_{t\in \mathcal{T}}\mathbf{x}[t]\mathbf{x}^{H}[t] \approx  \mathbb{E}\left\{ \mathbf{x}[t]\mathbf{x}^{H}[t] \right\} \\
	& = \sum\nolimits_{k\in \mathcal{K}}\mathbf{W}_{k} + \mathbf{R}_{0}\nonumber \succeq 0,
\end{align}
where $\mathbf{W}_{k} = \mathbf{w}_{k}\mathbf{w}_{k}^{H}$ with ${\rm rank}(\mathbf{W}_{k}) = 1$, and the statistical and sample covariance matrices are assumed to be approximately the same by considering a sufficiently large $T$. By letting $P_{\text{t}}$ denote the maximum transmit power at the BS, the total sample covariance matrix needs to satisfy the maximum transmit power constraint at the BS as $\mathrm{tr}\left(\mathbf{R}_{\mathbf{x}}\right) \leq P_{\text{t}}$. 
\subsection{Communication Performance Metric}
First, we consider the wireless communication from the BS to the CUs assisted by the IRS. Because the LoS link from the BS to each CU $k$ is assumed to be blocked, the received signal at CU $k$ through the BS-IRS-CU $k$ link at time symbol $t$ is given by
\begin{align}
	{y}_{k}[t] &= \mathbf{h}_{k}^{H}\mathbf{\Psi}\left(\mathbf{G}\mathbf{x}[t]+{\mathbf{z}_{1}}[t] \right)+{z}_{k}[t]\nonumber\\
	&=\underbrace{\mathbf{h}_{k}^{H}\mathbf{\Psi}\mathbf{G}\mathbf{w}_{k}s_{k}[t]}_{\text{User's desired signal}} +\underbrace{\sum\nolimits_{k' \in \mathcal{K} \backslash \{k\}} \!\mathbf{h}_{k}^{H}\mathbf{\Psi}\mathbf{G}\mathbf{w}_{k'}s_{k'}[t]}_{\text{Inter-user interference}}\nonumber\\
	& +  \underbrace{\mathbf{h}_{k}^{H}\mathbf{\Psi}\mathbf{G} \mathbf{s}_{0}[t]}_{\substack{\text{Dedicated sensing} \\ \text{signal interference}}}+ \underbrace{\mathbf{h}_{k}^{H}\mathbf{\Psi}{\mathbf{z}_{1}}[t]}_{\substack{\text{Reflected noise} \\ \text{ at the IRS}}} +{z}_{k}[t], \label{received_sig_CU}
\end{align}
where $\mathbf{\Psi} = \text{diag}(\bm{\psi})$, ${\mathbf{z}_{1}}[t] \sim \mathcal{CN}(\mathbf{0},\sigma_{\text{r}}^2\mathbf{I}_{N})$ denotes amplification noise introduced by the active IRS, and ${z}_{k}[t]\sim \mathcal{CN}(0,\sigma_{\text{u}}^2) $ represents the additive white Gaussian noise (AWGN) at each CU. Based on the received signal model in \eqref{received_sig_CU}, the received SINR at CU $k$ is given by 
\begin{align}
	\!\!\!\gamma_{k}  \!=\!\frac{ \left|\bar{\mathbf{h}}_{k}^{H}\mathbf{w}_{k}\right|^2 }{{\sum\nolimits_{k' \in \mathcal{K} \backslash \{k\}}\!\! \left|\bar{\mathbf{h}}_{k}^{H}\mathbf{w}_{k'}\right|^2 \! +\!\bar{\mathbf{h}}_{k}^{H} \mathbf{R}_{0}\bar{\mathbf{h}}_{k} \!+\! \sigma_{\text{r}}^2\mathbf{h}_{k}^{H}\mathbf{\Psi}\mathbf{\Psi}^{H}\mathbf{h}_{k} + \sigma_{\text{u}}^2}},\label{sinr_I}
\end{align}
where $\bar{\mathbf{h}}_{k}^{H} = \mathbf{h}_{k}^{H}\mathbf{\Psi}\mathbf{G}$.
\subsection{Sensing Performance Metric}
Next, we consider the extended target sensing task. Let $\mathbf{E} \in \mathbb{C}^{N \times N}$ denote the complete target response matrix of the IRS-target-IRS link. This matrix captures the cumulative effect of all scatterers within the extended target's volume. Notably, the IRS amplifies the signal twice during the round-trip propagation. First, the active IRS amplifies the transmit signal $\mathbf{x}[t]$ from the BS as
\begin{align}
	{\mathbf{x}_{1}}[t] =  \mathbf{\Psi\mathbf{G}}\mathbf{x}[t]  + \mathbf{\Psi}{\mathbf{z}_{1}}[t]. \label{IRS_ref_sig11}
\end{align}
Second, when the target reflects the signal back to the active IRS, the IRS again amplifies the echo signal as
\begin{align}
	{\mathbf{x}_{2}}[t] & = \mathbf{\Psi}\left(\mathbf{E}{\mathbf{x}_{1}}[t] + {\mathbf{z}_{2}}[t]\right)\nonumber                                                                                                                      \\
	& =  \mathbf{\Psi}\mathbf{E}\mathbf{\Psi}\mathbf{G}\mathbf{x}[t] + \mathbf{\Psi}\mathbf{E}\mathbf{\Psi}{\mathbf{z}_{1}}[t] + \mathbf{\Psi}{\mathbf{z}_{2}}[t],\label{IRS_ref_sig12}
\end{align}
where ${\mathbf{z}_{2}}[t]\sim \mathcal{CN}(\mathbf{0},\sigma_{\text{r}}^2\mathbf{I}_{N})$ is the AWGN induced by the active IRS. Let $P_{\text{s}}$ denote the maximum transmit power budget at the IRS. Therefore, the transmit power constraint at the active IRS is given by
\begin{align}
	& \mathbb{E}\left\{ \|{\mathbf{x}_{1}}[t]\|^2 + \|{\mathbf{x}_{2}}[t]\|^2 \right\} = \nonumber \\
	& \mathrm{tr}\left(\mathbf{\Psi}\mathbf{E}\mathbf{\Psi}\mathbf{G}\mathbf{R}_{\mathbf{x}}\mathbf{G}^{H}\mathbf{\Psi}^{H}\mathbf{E}_{l}^{H}\mathbf{\Psi}^{H}\right) + \mathrm{tr}\left( \mathbf{\Psi}\mathbf{G}\mathbf{R}_{\mathbf{x}}\mathbf{G}^{H}\mathbf{\Psi}^{H} \right) \nonumber                                                                                                 \\
	& + \sigma_{\text{r}}^2\mathrm{tr}\left(\mathbf{\Psi}\mathbf{E}\mathbf{\Psi}\mathbf{\Psi}^{H}\mathbf{E}^{H}\mathbf{\Psi}^{H}\right) + 2 \sigma_{\text{r}}^2 \mathrm{tr}\left(\mathbf{\Psi}\mathbf{\Psi}^{H}\right) \leq P_{\text{s}}.\label{IRS_power_cons1} 
\end{align}
Based on the signal model at the IRS in \eqref{IRS_ref_sig12}, the echo signal received by the BS from the BS-IRS-target-IRS-BS path at time symbol $t$ is given by
\begin{align} \label{ReflectedRadar}
	{\mathbf{y}}[t]\!\! =\!  \mathbf{G}^{T}\mathbf{\Psi}\mathbf{E}\mathbf{\Psi}\mathbf{G}\mathbf{x}[t]  \!\!+\! \mathbf{G}^{T}\mathbf{\Psi}\mathbf{E}\mathbf{\Psi}{\mathbf{z}_{1}}[t] \!\!+\! \mathbf{G}^{T}\mathbf{\Psi}{\mathbf{z}_{2}}[t] \!\!+\! {\mathbf{z}}[t],
\end{align}
where  $\mathbf{z}[t]\sim \mathcal{CN}(\mathbf{0},\sigma_{\text{b}}^2\mathbf{I}_{M}) $ denotes the AWGN at the BS. 
By concatenating $\mathbf{X}= [\mathbf{x}[1],\ldots,\mathbf{x}[T]]$, ${\mathbf{Y}} = [{\mathbf{y}}[1],\ldots,{\mathbf{y}}[T]]$, ${\mathbf{Z}}_{1} = [{\mathbf{z}}_{1}[1],\ldots,{\mathbf{z}}_{1}[T]]$, ${\mathbf{Z}}_{2} = [{\mathbf{z}}_{2}[1],\ldots,{\mathbf{z}}_{2}[T]]$, and ${\mathbf{Z}} = [{\mathbf{z}}[1],\ldots,{\mathbf{z}}[T]]$, we have
\begin{align}
	\!\!{\mathbf{Y}} & \!=\!  \mathbf{G}^{T}\mathbf{\Psi}\mathbf{E}\mathbf{\Psi}\mathbf{G}\mathbf{X}  \!+\! \mathbf{G}^{T}\mathbf{\Psi}\mathbf{E}\mathbf{\Psi}{\mathbf{Z}_{1}} \!\!+\! \mathbf{G}^{T}\mathbf{\Psi}{\mathbf{Z}_{2}} \!+\! {\mathbf{Z}}.\label{EchoSigAtBS1}
\end{align}
Accordingly, based on the received echo signal ${\mathbf{Y}}$ in \eqref{EchoSigAtBS1}, the BS needs to estimate the complete target response matrix $\mathbf{E}$ for the extended target. It is worthy to note that the active IRS brings the additional amplification noise terms in \eqref{EchoSigAtBS1} which affects the estimation CRB of target sensing. In the following, we derive the estimation CRB for characterizing the sensing performance. 

By vectorizing the received echo signal $\mathbf{Y}$ in \eqref{EchoSigAtBS1}, we have
\begin{align}
	{\mathbf{y}} & = \mathrm{vec}({\mathbf{Y}})= {\bm{\eta}} +{\mathbf{w}},\label{vec_y_l_bs}
\end{align}
where ${\bm{\eta}} =\mathrm{vec}\left(\mathbf{G}^{T}\mathbf{\Psi}{\mathbf{E}}\mathbf{\Psi}\mathbf{G}\mathbf{X}\right) = \left(\mathbf{X}^{T}\mathbf{G}^{T}\mathbf{\Psi} \otimes \mathbf{G}^{T}\mathbf{\Psi}\right){\bm{\xi}}$ with $\bm{\xi} = \mathrm{vec}\left(\mathbf{E}\right)$ and ${\mathbf{w}} = \mathrm{vec}(\mathbf{G}^{T}\mathbf{\Psi}{\mathbf{Z}_{2}} + {\mathbf{Z}}) = (\mathbf{I}_{T} \otimes \mathbf{G}^{T}\mathbf{\Psi}) \mathrm{vec}( {\mathbf{Z}_{2}}) + \mathrm{vec}( {\mathbf{Z}})$. Note that in \eqref{vec_y_l_bs}, we ignore the noise term of ${\mathbf{z}_{1}}$ due to the fact that it suffers from the triple-reflection path loss induced by the IRS-target-IRS-BS link, whose power can be neglected \cite{zhu2024cramer}. Let $\mathbf{p} =[a_1,\ldots,a_N]^{T}$, $\bm{\phi} = [e^{\jmath\rho_{1}},\ldots,e^{\jmath\rho_{N}}]^{T}$, $\mathbf{P} = \mathrm{diag}(\mathbf{p})$, and $\mathbf{\Phi} = \mathrm{diag}(\bm{\phi})$. We thus have  $\mathbf{\Psi } = \mathbf{P}\mathbf{\Phi}$.
Then, the mean vector and covariance matrix of ${\mathbf{y}}$ are given by ${\bm{\eta}}$ and $\mathbf{R}_{{\mathbf{y}}} =\mathbf{I}_{T} \otimes \mathbf{R}_{{\mathbf{w}}}$, respectively, where $\mathbf{R}_{{\mathbf{w}}} = \sigma_{\text{r}}^{2}  \mathbf{G}^{T}\mathbf{\Psi}\mathbf{\Psi}^{H}\mathbf{G}^{*}+\sigma_{\text{b}}^{2}\mathbf{I}_{{M}} = \sigma_{\text{r}}^{2}  \mathbf{G}^{T}\mathbf{P}^{2}\mathbf{G}^{*}+\sigma_{\text{b}}^{2}\mathbf{I}_{{M}} $. According to the definition of CRB, the CRB for estimating parameter vector $\bm{\xi}$ is given by $\mathrm{CRB}_{\mathbf{E}}(\mathbf{R}_{\mathbf{x}},\mathbf{\Psi})=\mathrm{tr}(\mathbf{F}^{-1})$, where ${\mathbf{F}}$ denotes the Fisher information matrix (FIM) with respect to $\bm{\xi}$. According to estimation theory, the $(p,q)$-th element of ${\mathbf{F}}$ is given by \cite{kay1993fundamentals}
\begin{align}
	  &[{\mathbf{F}}]_{p,q} = \nonumber\\
	  &\mathrm{tr}\left(\mathbf{R}_{{\mathbf{y}}}^{-1}\frac{\partial \mathbf{R}_{{\mathbf{y}}}}{\partial [{\bm{\xi}}]_{p}}\mathbf{R}_{{\mathbf{y}}}^{-1}\frac{\partial \mathbf{R}_{{\mathbf{y}}}}{\partial [{\bm{\xi}}]_{q}}\right) + 2\Re\left(\frac{\partial {\bm{\eta}}^{H}}{\partial [{\bm{\xi}}]_{p} } \mathbf{R}_{{\mathbf{y}}}^{-1} \frac{\partial {\bm{\eta}}}{\partial [{\bm{\xi}}]_{q} }\right). \label{FIM_def_bs}
\end{align}
Based on \eqref{FIM_def_bs}, we have the following proposition.
\begin{prop}\label{prop_FIM_bs}
	The FIM for estimating ${\mathbf{F}}$ is given by
	\begin{align}
	\mathbf{F} = \left[\begin{array}{cc}
		\mathbf{F}_{\Re\{{\bm{\xi}}\},\Re\{{\bm{\xi}}\}}&\mathbf{F}_{\Re\{{\bm{\xi}}\},\Im\{{\bm{\xi}}\}}\\
		\mathbf{F}_{\Im\{{\bm{\xi}}\},\Re\{{\bm{\xi}}\}}& \mathbf{F}_{\Im\{{\bm{\xi}}\},\Im\{{\bm{\xi}}\}}
	\end{array}\right], \label{FIM_extended}
\end{align}
where
	\begin{align}
		&\!\!\!\mathbf{F}_{\Re\{{\bm{\xi}}\},\Re\{{\bm{\xi}}\}} =\mathbf{F}_{\Im\{{\bm{\xi}}\},\Im\{{\bm{\xi}}\}} \nonumber\\
		&\!\!\!=\!2\Re \! \left\{ \!{\left( {{{\mathbf{\Psi }}^H}{\mathbf{G}^*}{{\bf{X}}^*} \!\otimes\! {{\mathbf{\Psi }}^H}{\mathbf{G}^*}} \right)\!{\bf{R}}_{\bf{y}}^{ - 1}\!\left( {{{\bf{X}}^T}{\mathbf{G}^T}{\mathbf{\Psi }} \! \otimes \! {\mathbf{G}^T}{\mathbf{\Psi }}} \right)} \! \right\}\!, \label{FIM_Imh_Imh}\\
		&\!\!\!\mathbf{F}_{\Im\{{\bm{\xi}}\},\Re\{{\bm{\xi}}\}} = -\mathbf{F}_{\Re\{{\bm{\xi}}\},\Im\{{\bm{\xi}}\}}\nonumber\\
		&\!\!\!=\! 2\Im \! \left\{\! {\left( {{{\mathbf{\Psi }}^H}{\mathbf{G}^*}{{\bf{X}}^*} \!\otimes\! {{\mathbf{\Psi }}^H}{\mathbf{G}^*}} \right)\!{\bf{R}}_{\bf{y}}^{ - 1}\!\left( {{{\bf{X}}^T}{\mathbf{G}^T}{\mathbf{\Psi }}\! \otimes \! {\mathbf{G}^T}{\mathbf{\Psi }}} \right)}\! \right\}\!.\label{FIM_Reh_Imh}
		\end{align}
\end{prop}
\begin{IEEEproof}
	See Appendix \ref{prop_FIM1_proof}.
\end{IEEEproof}
Subsequently, the estimation CRB is given by 
{\small
\begin{align}
	&\mathrm{CRB}_{\mathbf{E}}(\mathbf{R}_{\mathbf{x}},\mathbf{\Psi})=\mathrm{tr}(\mathbf{F}^{-1})= \nonumber\\
	&=\mathrm{tr}\left(\left( {{\mathbf{\Psi }}^H}{\mathbf{G}^*}{{\bf{X}}^*} {{\bf{X}}^T}{\mathbf{G}^T}{\mathbf{\Psi }} \otimes {{\mathbf{\Psi }}^H}{\mathbf{G}^*}{\bf{R}}_{\bf{w}}^{ - 1} {\mathbf{G}^T}{\mathbf{\Psi }} \right)^{-1}\right)\nonumber\\
	&=\frac{1}{T}\mathrm{tr}\left(\left( {{\mathbf{\Psi }}^H}{\mathbf{G}^*} \mathbf{R}_{\mathbf{x}}^{T} {\mathbf{G}^T}{\mathbf{\Psi }} \right)^{-1} \otimes \left( {{\mathbf{\Psi }}^H}{\mathbf{G}^*}{\bf{R}}_{\bf{w}}^{ - 1} {\mathbf{G}^T}{\mathbf{\Psi }} \right)^{-1} \right)\nonumber\\
	&= \frac{1}{T}\mathrm{tr}\left(\left({\mathbf{G}} \mathbf{R}_{\mathbf{x}} {\mathbf{G}^H} \right)^{\!\!-1}\!{\mathbf{P}}^{-2}\right) \mathrm{tr}\left( \left( {\mathbf{G}}{\bf{R}}_{\bf{w}}^{ - 1} {\mathbf{G}^H} \right)^{\!\!-1}\!{\mathbf{P}}^{-2} \right). \label{CRB_exp}
\end{align}}
\begin{rem}
	Recall that the CRB under the passive IRS deployment is independent of $\mathbf{\Psi}$ \cite{song2023intelligent}. In contrast, the CRB in \eqref{CRB_exp} for an extended target depends on the reflective beamforming $\mathbf{\Psi}$ when an active IRS is deployed. More specifically, it is shown that the CRB depends solely on the amplification gain $\mathbf{P}$ at the active IRS. Furthermore,  the condition $M\geq \mathrm{rank}\left(\mathbf{R}_{\mathbf{x}}\right)\geq N \geq \mathrm{rank}\left(\mathbf{G}\right)$ must hold to guarantee the CRB in \eqref{CRB_exp} to be bounded, such that the target response matrix $\mathbf{E}$ is estimatable. 
\end{rem}

\subsection{Problem Formulation}
Our goal is to jointly design the transmit beamforming $\{\mathbf{w}_{k},\mathbf{R}_{0}\}$ at the BS and the reflective beamforming $\mathbf{\Psi}$ at the IRS for improving the sensing performance of the extended target. \textcolor{black}{Specifically, we formulate the following optimization problem (P1), by leveraging the closed-form CRB derived in \eqref{CRB_exp}.}
\begin{subequations}
	\begin{eqnarray}
		(\text{P1}):&\!\!\!\!\!\!{\mathop {\min}\limits_{\mathbf{\Psi},\{\mathbf{w}_{k},\mathbf{R}_{0}\}} }\!\!\!\!\!\!& \mathrm{CRB}_{\mathbf{E}}(\mathbf{R}_{\mathbf{x}},\mathbf{\Psi}) \nonumber\\
		&\!\!\!\!\!\!\text{s.t.}\!\!\!\!\!\! & \gamma_{k} \geq \Gamma_{k}, \forall k \in \mathcal{K}, \label{P1_I_cons1}\\
		&&\eqref{IRS_power_cons1}, \label{P1_I_cons2}\\
		&& \mathrm{tr}\left(\mathbf{R}_{\mathbf{x}}\right) \leq P_{\text{t}}, \label{P1_I_cons3}\\	
		&&\mathbf{R}_{\mathbf{x}} \succeq \mathbf{0}, \label{P1_I_cons4}\\
		&& \left|[\mathbf{\Psi}]_{n,n}\right| \leq a_{\text{max}}, \forall n \in \mathcal{N}. \label{P1_I_cons5}
	\end{eqnarray}
\end{subequations}
In problem (P1), \eqref{P1_I_cons1} denotes the SINR requirements at the CUs, \eqref{P1_I_cons2} denotes the transmit power constraints at the IRS,  \eqref{P1_I_cons3} denotes the transmit power constraint at the BS, \eqref{P1_I_cons4} denotes the semi-definite constraint regarding the sample covariance matrix of the transmit signal, and \eqref{P1_I_cons2} denotes the maximum power amplification gain constraints at the IRS, respectively. Problem (P1) is highly non-convex due to the non-convexity of both the objective function and the constraints in \eqref{P1_I_cons1} and \eqref{P1_I_cons2}. To address this issue, we adopt the AO approach, where the transmit beamforming $\{\mathbf{w}_{k},\mathbf{R}_{0}\}$ at the BS and the reflective beamforming $\mathbf{\Psi}$ at the active IRS are optimized in an iterative manner.

\section{Joint Beamforming Design for Sensing-Only Scenario}
In this section, we consider the sensing-only scenario and propose an efficient joint beamforming design through AO method. As there is sensing-only task considered, we ignore the SINR constraints in \eqref{P1_I_cons1}, such that problem (P1) is reduced to the following problem:
	\begin{eqnarray}
			&\!\!\!(\text{P2}):\!\!\!&  \nonumber\\
			&\!\!\!\!\!\!{\mathop {\min}\limits_{\mathbf{\Psi},\{\mathbf{R}_{x}\}} }\!\!\!\!\!\!& \mathrm{tr}\left(\left({\mathbf{G}} \mathbf{R}_{\mathbf{x}} {\mathbf{G}^H} \right)^{\!\!-1}\!{\mathbf{P}}^{-2} \right) \mathrm{tr}\left( \left( {\mathbf{G}}{\bf{R}}_{\bf{w}}^{ - 1} {\mathbf{G}^H} \right)^{\!\!-1}\!{\mathbf{P}}^{-2} \right) \nonumber\\
		&\!\!\!\!\!\!\text{s.t.}\!\!\!\!\!\! &\eqref{P1_I_cons2}-\eqref{P1_I_cons4}.\nonumber
	\end{eqnarray}
Note that problem (P2) is non-convex due to the couple of optimization variables and non-convex constrint \eqref{P1_I_cons2}. In the following, we propose an AO optimization to solve the problem by alternately optimizing transmit signal beamforming $\mathbf{R}_{x}$ and reflective beamforming $\mathbf{\Psi}$.
\subsection{Optimal Transmit Signal Covariance with Given $\mathbf{\Psi}$} 
Under a given reflective beamforming matrix $\mathbf{\Psi}$, the optimization of transmit beamforming is reformulated as 
	\begin{eqnarray}
			\!\!\!(\text{P3}):\!\!\!& {\mathop {\min}\limits_{\mathbf{R}_{\mathbf{x}}} }& \mathrm{tr}\left(\left({\mathbf{G}} \mathbf{R}_{\mathbf{x}} {\mathbf{G}^H} \right)^{\!\!-1}\!{\mathbf{P}}^{-2} \right)  \nonumber\\
		&\text{s.t.} & \eqref{P1_I_cons2}-\eqref{P1_I_cons4}.\nonumber
	\end{eqnarray}
Note that problem (P3) is a convex problem that can be solved by existing solvers such as CVX \cite{grant2014cvx}. Nevertheless, under certain conditions, the optimal closed-form solution to problem (P3) can be obtained to provide insights and reduce computational complexity.

First, we have $\mathrm{tr}\left(\left({\mathbf{G}} \mathbf{R}_{\mathbf{x}} {\mathbf{G}^H} \right)^{\!-1}\!{\mathbf{P}}^{-2} \right) =  \mathrm{tr}\left(\left({\mathbf{P}}{\mathbf{G}} \mathbf{R}_{\mathbf{x}} {\mathbf{G}^H}{\mathbf{P}} \right)^{-1}\right)$ for the objective function  and then apply the singular value decomposition (SVD) to ${\mathbf{P}}{\mathbf{G}}$ as ${\mathbf{P}}{\mathbf{G}}=\mathbf{U}_{1}\bm{\Sigma}\mathbf{U}_{2}^{H}$, where $\bm{\Sigma} = [\bm{\Sigma}_{1}, \mathbf{0}]$ and $\bm{\Sigma}_{1} = \mathrm{diag}(\sigma_{1},\ldots,\sigma_{N})$. Then, we have 
\begin{align}
	&\mathrm{tr}\left(\left({\mathbf{P}}{\mathbf{G}} \mathbf{R}_{\mathbf{x}} {\mathbf{G}^H}{\mathbf{P}} \right)^{-1}\right)= \mathrm{tr}\left(\left(\bm{\Sigma}^{2}\bar{\mathbf{R}}_{\mathbf{x}} \right)^{-1}\right),
\end{align} 
where $\bar{\mathbf{R}}_{\mathbf{x}} = \mathbf{U}_{2}^{H}\mathbf{R}_{\mathbf{x}}\mathbf{U}_{2} \in \mathbb{C}^{M \times M}$. Then, the transmit power constraint at the BS in \eqref{P1_I_cons3} is rewritten as
\begin{align}
	\mathrm{tr}\left(\mathbf{U}_{2}\bar{\mathbf{R}}_{\mathbf{x}}\mathbf{U}_{2}^{H}\right) = \mathrm{tr}\left(\bar{\mathbf{R}}_{\mathbf{x}}\right) \leq P_{\text{t}}, \label{P1_I_cons3_eq}
\end{align}
and the transmit power constraint at the IRS in \eqref{P1_I_cons2} is rewritten as 
\begin{align}
	& \mathrm{tr}\left(\mathbf{\Psi}\mathbf{E}\mathbf{\Psi}\mathbf{G}\mathbf{U}_{2}\bar{\mathbf{R}}_{\mathbf{x}}\mathbf{U}_{2}^{H}\mathbf{G}^{H}\mathbf{\Psi}^{H}\mathbf{E}^{H}\mathbf{\Psi}^{H}\right)+ 2 \sigma_{\text{r}}^2 \mathrm{tr}\left(\mathbf{\Psi}\mathbf{\Psi}^{H}\right)\!+\! \nonumber                                                                                                 \\
	& \sigma_{\text{r}}^2\mathrm{tr}\!\left(\mathbf{\Psi}\mathbf{E}\mathbf{\Psi}\mathbf{\Psi}^{H}\mathbf{E}^{H}\mathbf{\Psi}^{H}\right)\!\!+\! \mathrm{tr}\!\left(\! \mathbf{\Psi}\mathbf{G}\mathbf{U}_{2}\bar{\mathbf{R}}_{\mathbf{x}}\mathbf{U}_{2}^{H}\mathbf{G}^{H}\mathbf{\Psi}^{H} \!\right) \leq P_{\text{s}} .\label{IRS_power_cons1_eq1} 
\end{align}
Note that constraint \eqref{IRS_power_cons1_eq1} is further equivalent to 
\begin{align}
	 \mathrm{tr}\left(\mathbf{A}\bar{\mathbf{R}}_{\mathbf{x}}\right) \leq \bar{P}_{\text{s}},\label{IRS_power_cons1_eq2}
\end{align}
where $\bar{P}_{\text{s}}=P_{\text{s}}-\sigma_{\text{r}}^2\mathrm{tr}\left(\mathbf{P}^{2}\mathbf{E}\mathbf{P}^{2}\mathbf{E}^{H}\right)-2 \sigma_{\text{r}}^2 \mathrm{tr}\left(\mathbf{P}^{2}\right)$ and $\mathbf{A}=\mathbf{U}_{2}^{H}\mathbf{G}^{H}\mathbf{\Psi}^{H}\left(\mathbf{E}^{H}\mathbf{P}^{2}\mathbf{E}+\mathbf{I}_{N}\right)\mathbf{\Psi}\mathbf{G}\mathbf{U}_{2}$. According to the trace inequality \cite{yang2000matrix}, we have 
the following lemma.
\begin{lem} \label{trace_inq_lem}
	If $\frac{\bar{P}_{\text{s}}}{P_{\text{t}}}\mathbf{I}_{M}-\mathbf{A} \succeq \mathbf{0}$ or $\frac{\bar{P}_{\text{s}}}{P_{\text{t}}} \geq  \mathrm{tr}\left(\mathbf{A}\right) $ is satisfied, then the constraint in \eqref{P1_I_cons3_eq} implies that in \eqref{IRS_power_cons1_eq2}.
\end{lem}
\begin{IEEEproof}
	First, when $\frac{\bar{P}_{\text{s}}}{P_{\text{t}}}\mathbf{I}_{M}-\mathbf{A} \succeq \mathbf{0}$, it follows based on the trace inequality that $\mathrm{tr}\left(\left(\frac{\bar{P}_{\text{s}}}{P_{\text{t}}}\mathbf{I}_{M}-\mathbf{A}\right)\bar{\mathbf{R}}_{\mathbf{x}}\right) \geq 0$. Thus, $\mathrm{tr}\left(\frac{\bar{P}_{\text{s}}}{P_{\text{t}}}\bar{\mathbf{R}}_{\mathbf{x}}\right) \geq \mathrm{tr}\left(\mathbf{A}\bar{\mathbf{R}}_{\mathbf{x}}\right)$. Furthermore, as the constraint in \eqref{P1_I_cons3_eq} holds, we have $\mathrm{tr}\left(\mathbf{A}\bar{\mathbf{R}}_{\mathbf{x}}\right) \leq \frac{\bar{P}_{\text{s}}}{P_{\text{t}}}\mathrm{tr}\left(\bar{\mathbf{R}}_{\mathbf{x}}\right) \leq \bar{P}_{\text{s}} $. 
	
	On the other hand, when $\frac{\bar{P}_{\text{s}}}{P_{\text{t}}} \geq  \mathrm{tr}\left(\mathbf{A}\right) $, we have $\mathrm{tr}\left(\mathbf{A}\bar{\mathbf{R}}_{\mathbf{x}}\right) \leq \mathrm{tr}\left(\mathbf{A}\right)\mathrm{tr}\left(\bar{\mathbf{R}}_{\mathbf{x}}\right) \leq \frac{\bar{P}_{\text{s}}}{P_{\text{t}}} \mathrm{tr}\left(\bar{\mathbf{R}}_{\mathbf{x}}\right)$. As the constraint in \eqref{P1_I_cons3_eq} holds, we have $\mathrm{tr}\left(\mathbf{A}\bar{\mathbf{R}}_{\mathbf{x}}\right) \leq \frac{\bar{P}_{\text{s}}}{P_{\text{t}}}\mathrm{tr}\left(\bar{\mathbf{R}}_{\mathbf{x}}\right) \leq \bar{P}_{\text{s}} $.
	As a result, Lemma \ref{trace_inq_lem} is proved.
\end{IEEEproof}
%

Based on Lemma \ref{trace_inq_lem}, we then propose a scheme to obtain the optimal closed-form solution to problem (P3) in the case when  $\frac{\bar{P}_{\text{s}}}{P_{\text{t}}}\mathbf{I}_{M}-\mathbf{A} \succeq \mathbf{0}$ or $\frac{\bar{P}_{\text{s}}}{P_{\text{t}}} \geq  \mathrm{tr}\left(\mathbf{A}\right) $ is satisfied. Accordingly, we denote the matrix $\bar{\mathbf{R}}_{\mathbf{x}}$ as
\begin{align}
	\bar{\mathbf{R}}_{\mathbf{x}} = 	\left[\begin{array}{cc}
	\bar{\mathbf{R}}_{\mathbf{x},1}&\bar{\mathbf{R}}_{\mathbf{x},2}\\
	\bar{\mathbf{R}}_{\mathbf{x},2}^{H}&\bar{\mathbf{R}}_{\mathbf{x},3}
	\end{array}\right],\label{R_x_dec}
\end{align}
where $\bar{\mathbf{R}}_{\mathbf{x},1}\in \mathbb{C}^{N\times N}$. Thus, we have $\mathrm{tr}\left(\left(\bm{\Sigma}^{2}\bar{\mathbf{R}}_{\mathbf{x}} \right)^{-1}\right) = \mathrm{tr}\left(\left(\bm{\Sigma}_{1}^{2}\bar{\mathbf{R}}_{\mathbf{x},1} \right)^{-1}\right)$ and problem (P3) is reformulated as 
	\begin{eqnarray}
			\!\!\!(\text{P3.1}):\!\!\!& {\mathop {\min}\limits_{\bar{\mathbf{R}}_{\mathbf{x}}} }& \mathrm{tr}\left(\left(\bm{\Sigma}_{1}^{2}\bar{\mathbf{R}}_{\mathbf{x},1} \right)^{-1}\right)  \nonumber\\
		&\text{s.t.} & \eqref{P1_I_cons4}, \eqref{P1_I_cons3_eq}, \text{and }  \eqref{R_x_dec}.\nonumber
	\end{eqnarray}
According to \cite[Lemma 1]{yang2007mimo}, we have 
\begin{align}
	\mathrm{tr}\left(\left(\bm{\Sigma}_{1}^{2}\bar{\mathbf{R}}_{\mathbf{x},1} \right)^{-1}\right) \geq \sum\nolimits_{i=1}^{N} \frac{1}{\sigma_{i}^{2}r_{i}},
\end{align}
where $r_{i}=[\bar{\mathbf{R}}_{\mathbf{x},1}]_{i,i}$. Consequently, we obtain the following proposition.
\begin{prop}\label{prop1}
	At the optimal solution to problem (P3.1), $\bar{\mathbf{R}}_{\mathbf{x},2}$ and $\bar{\mathbf{R}}_{\mathbf{x},3}$ are zero matrices, and $\bar{\mathbf{R}}_{\mathbf{x},1} = \mathrm{diag}\left(r_{1},\ldots,r_{N}\right)$ is diagonal, where $r_{i}\geq 0, \forall i \in \mathcal{N}$.
\end{prop}
\begin{IEEEproof}
	The proof is similar to \cite[Proposition 3]{song2023intelligent}, and thus is omitted for brevity.
\end{IEEEproof}

According to Proposition \ref{prop1}, problem (P3.1) is reformulated as
\begin{subequations}
	\begin{eqnarray}
			\!\!\!(\text{P3.2}):\!\!\!& {\mathop {\min}\limits_{ \{r_{i}\} } }& \sum\nolimits_{i=1}^{N} \frac{1}{\sigma_{i}^{2}r_{i}}  \nonumber\\
		&\text{s.t.} & \sum\nolimits_{i=1}^{N} r_{i} \leq {P}_{\text{t}},\\
		&& r_{i} \geq 0, \forall i \in \mathcal{N}.
	\end{eqnarray}
\end{subequations}
Based on the Karush-Kuhn-Tucker (KKT) conditions, the optimal solution to problem (P3.2) is given by the following proposition.
\begin{prop}\label{prop3}
 The optimal solution to problem (P3.2) is given by $r_{i}^{\star} = \frac{\sigma_{i}^{-1}{P}_{\text{t}}}{\sum\nolimits_{i=1}^{N} \sigma_{i}^{-1}}, \forall i \in \mathcal{N}$.
\end{prop}
\begin{IEEEproof}
	The proof is similar to \cite[Proposition 4]{song2023intelligent}, and thus is omitted for brevity.
\end{IEEEproof}
Based on Proposition \ref{prop3}, the optimal solution to problem (P3.1) is given by 
\begin{align}
	\bar{\mathbf{R}}_{\mathbf{x}}^{\star} = 	\left[\begin{array}{cc}
		\bm{\Sigma}_{1}^{-1}&\mathbf{0}\\
		\mathbf{0}&\mathbf{0}
	\end{array}\right].
\end{align}
Furthermore, recall that $\bar{\mathbf{R}}_{\mathbf{x}} = \mathbf{U}_{2}^{H}\mathbf{R}_{\mathbf{x}}\mathbf{U}_{2}$ based on which the optimal solution to problem (P3) under the conditions $\frac{\bar{P}_{\text{s}}}{P_{\text{t}}}\mathbf{I}_{M}-\mathbf{A} \succeq \mathbf{0}$ or $\frac{\bar{P}_{\text{s}}}{P_{\text{t}}} \geq  \mathrm{tr}\left(\mathbf{A}\right) $ is given by 
\begin{align}
		{\mathbf{R}}_{\mathbf{x}}^{\star} = 	\mathbf{U}_{2}^{H}\bar{\mathbf{R}}_{\mathbf{x}}^{\star}\mathbf{U}_{2}.\label{R_x_opt}
\end{align}

\begin{rem}
	The conditions $\frac{\bar{P}_{\text{s}}}{P_{\text{t}}}\mathbf{I}_{M}-\mathbf{A} \succeq \mathbf{0}$ and $\frac{\bar{P}_{\text{s}}}{P_{\text{t}}} \geq  \mathrm{tr}\left(\mathbf{A}\right)$ suggest that a small maximum transmit power $P_{\text{t}}$ at the BS and a large  $\bar{P}_{\text{s}}$ increase the likelihood that problem (P3) has a closed-form solution. These conditions are sufficient and can help determine whether to use a convex optimizer for solving the transmit beamforming optimization or to directly obtain the optimal closed-form transmit beamforming solution using \eqref{R_x_opt}.
\end{rem}

\subsection{Reflective Beamforming Optimization with Given $\mathbf{R}_{\mathbf{x}}$}
Under given transmit beamforming $\mathbf{R}_{\mathbf{x}}$, the optimization of reflective beamforming is reformulated as
	\begin{eqnarray}
			&\!\!\!(\text{P4}):\!\!\!&  \nonumber\\
			&\!\!\!\!\!\!{\mathop {\min}\limits_{\mathbf{\Psi} } }\!\!\!\!\!\!& \mathrm{tr}\left(\left({\mathbf{G}} \mathbf{R}_{\mathbf{x}} {\mathbf{G}^H} \right)^{\!\!-1}\!{\mathbf{P}}^{-2} \right) \mathrm{tr}\left( \left( {\mathbf{G}}{\bf{R}}_{\bf{w}}^{ - 1} {\mathbf{G}^H} \right)^{\!\!-1}\!{\mathbf{P}}^{-2} \right) \nonumber\\
		&\!\!\!\!\!\!\text{s.t.}\!\!\!\!\!\! &\eqref{P1_I_cons2}~\text{and}~\eqref{P1_I_cons5}.\nonumber
	\end{eqnarray}
Based on the definition $\mathbf{\Psi } = \mathbf{P}\mathbf{\Phi}$, the constraints in \eqref{P1_I_cons2} and \eqref{P1_I_cons5} are rewritten as 
\begin{align}
	& \mathrm{tr}\left(\mathbf{P}^{2}\mathbf{E}\mathbf{P}\mathbf{\Phi}\mathbf{G}\mathbf{R}_{\mathbf{x}}\mathbf{G}^{H}\mathbf{\Phi}^{H}\mathbf{P}\mathbf{E}^{H}\right)+ 2 \sigma_{\text{r}}^2 \mathrm{tr}\left(\mathbf{P}^{2}\right) \nonumber                                                                                                 \\
	& + \sigma_{\text{r}}^2\mathrm{tr}\left(\mathbf{P}^{2}\mathbf{E}\mathbf{P}^{2}\mathbf{E}^{H}\right)\!\!+\! \mathrm{tr}\left( \mathbf{P}^{2}\mathbf{G}\mathbf{R}_{\mathbf{x}}\mathbf{G}^{H} \right) \leq P_{\text{s}},\label{IRS_power_cons1_eq3}
\end{align}
and 
\begin{align}
\left\{
\begin{array}{ll}
	0 \leq [\mathbf{P}]_{i,i} \leq a_{\text{max}},& \forall i \in \mathcal{N},\\
	 \left[\mathbf{P}\right]_{i,j} = 0, & \forall i \neq j,
\end{array}
	\right.\label{IRS_power_cons5_eq}
\end{align}
respectively. 
Thus, problem (P4) is equivalent to 
\begin{subequations}
	\begin{eqnarray}
			&\!\!\!(\text{P4.1}):\!\!\!&  \nonumber\\
			&\!\!\!\!\!\!{\mathop {\min}\limits_{\mathbf{P}, \mathbf{\Phi} } }\!\!\!\!\!\!& \mathrm{tr}\left(\left({\mathbf{G}} \mathbf{R}_{\mathbf{x}} {\mathbf{G}^H} \right)^{\!\!-1}\!{\mathbf{P}}^{-2} \right) \mathrm{tr}\left( \left( {\mathbf{G}}{\bf{R}}_{\bf{w}}^{ - 1} {\mathbf{G}^H} \right)^{\!\!-1}\!{\mathbf{P}}^{-2} \right) \nonumber\\
		&\!\!\!\!\!\!\text{s.t.}\!\!\!\!\!\! & [\mathbf{\Phi}]_{i,i} = 1, \forall i \in \mathcal{N},\label{Phi_mod_cons} \\
		&& [\mathbf{\Phi}] \succeq \mathbf{0}, \label{Phi_semi_cons} \\
		&&\eqref{IRS_power_cons1_eq3}~\text{and}~\eqref{IRS_power_cons5_eq}.\nonumber
	\end{eqnarray}
\end{subequations}
Problem (P4.1) is non-convex due to the non-convexity of the objective function and the constraints in \eqref{IRS_power_cons1_eq3} and \eqref{IRS_power_cons5_eq}. In the following, we will propose an alternating optimization algorithm to efficiently solve problem (P4.1).
\subsubsection{Optimization of $\mathbf{\Phi}$}
Note that in problem (P4.1), the objective function and the constraint in \eqref{IRS_power_cons5_eq} are independent of $\mathbf{\Phi}$. \textcolor{black}{Based on this observation, the optimal $\mathbf{\Phi}$ should make the feasible region of $\mathbf{P}$ as large as possible and this relates to the constraint in \eqref{IRS_power_cons1_eq3}. It is observed that optimizing $\mathbf{\Phi}$ to minimize the left-hand term in \eqref{IRS_power_cons1_eq3} is able to enlarge the feasible region of $\mathbf{P}$.} Thus, we optimize $\mathbf{\Phi}$ via solving the following problem:
	\begin{eqnarray}
			(\text{P4.2}):&{\mathop {\min}\limits_{\mathbf{\Phi} } }&  \mathrm{tr}\left(\mathbf{P}^{2}\mathbf{E}\mathbf{P}\mathbf{\Phi}\mathbf{G}\mathbf{R}_{\mathbf{x}}\mathbf{G}^{H}\mathbf{\Phi}^{H}\mathbf{P}\mathbf{E}^{H}\right) \nonumber\\
		&\text{s.t.} & \eqref{Phi_mod_cons}~\text{and}~\eqref{Phi_semi_cons}.\nonumber
	\end{eqnarray}
Based on the property of vectorization operator \cite[Chapter 10.2.2]{petersen2008matrix}, we have  $\mathrm{tr}\left(\mathbf{P}^{2}\mathbf{E}\mathbf{P}\mathbf{\Phi}\mathbf{G}\mathbf{R}_{\mathbf{x}}\mathbf{G}^{H}\mathbf{\Phi}^{H}\mathbf{P}\mathbf{E}^{H}\right) = \mathrm{vec}^{H}\left(\mathbf{\Phi}\right)\left(\mathbf{C}^{T}\otimes \mathbf{B} \right) \mathrm{vec}\left(\mathbf{\Phi}\right)$, where $\mathbf{B} = \mathbf{P}\mathbf{E}^{H}\mathbf{P}^{2}\mathbf{E}\mathbf{P}$ and $\mathbf{C} = \mathbf{G}\mathbf{R}_{\mathbf{x}}\mathbf{G}^{H}$.  Consequently, problem (P4.2) is equivalent to 
	\begin{eqnarray}
			(\text{P4.3}):&{\mathop {\min}\limits_{\mathbf{\Phi} } }&  \mathrm{vec}^{H}\left(\mathbf{\Phi}\right)\left(\mathbf{C}^{T}\otimes \mathbf{B} \right) \mathrm{vec}\left(\mathbf{\Phi}\right) \nonumber\\
		&\text{s.t.} & \eqref{Phi_mod_cons}~\text{and}~\eqref{Phi_semi_cons}.\nonumber
	\end{eqnarray}
We define $\mathbf{v} \triangleq \mathrm{vec}\left(\mathbf{\Phi}\right)$ and $\mathbf{V} \triangleq \mathbf{v}\mathbf{v}^{H}$ with $\mathrm{rank}\left(\mathbf{V}\right) = 1$. According to \eqref{Phi_mod_cons} and \eqref{Phi_semi_cons}, we have 
\begin{align}
	&[\mathbf{V}]_{n,n} = \left\{\begin{array}{l}
		1, \quad n \in \{(i-1)N+i|i\in \mathcal{N}\}, \\
		0, \quad \text{otherwise},
	\end{array} \right. \label{V_mod_cons}\\
	&\mathbf{V} \succeq \mathbf{0}.\label{V_semi_cons}
\end{align}
Based on the SDR technique, problem (P4.3) is relaxed as problem (SDR4.3) by dropping the rank-one constraint.
	\begin{eqnarray}
			(\text{SDR4.3}):&{\mathop {\min}\limits_{\mathbf{V} } }&  
			\mathrm{tr}\left(\left(\mathbf{C}^{T}\otimes \mathbf{B} \right) \mathbf{V}\right)  \nonumber\\
		&\text{s.t.} & \eqref{V_mod_cons}~\text{and}~\eqref{V_semi_cons}.\nonumber
	\end{eqnarray}
Note that problem (SDR4.3) is a convex semi-definite program (SDP), which can be optimally solved by existing solvers like CVX. Let $\mathbf{V}^{\star}$ denote the optimal solution to problem (SDR4.3), which may not be of rank-one. To tackle this issue, we further adopt the Gaussian randomization to construct a high-quality rank-one solution to problem (SDR4.3) based on the obtained $\mathbf{V}^{\star}$. Specifically, we first generate a number of generated randomly vectors ${\mathbf{r}} \sim \mathcal{CN}\left(\mathbf{0},\mathbf{I}_{N^{2}}\right)$, and then construct a number of candidate rank-one solutions as 
\begin{align}\label{GR_construct}
	\mathbf{v} = \bm{\zeta} \circ e^{\jmath\mathrm{arg}\left\{\left(\mathbf{V}^{\star}\right)^{\frac{1}{2}}\mathbf{r}\right\}},
\end{align}	
where $ \bm{\zeta} $ is an $N^{2}$-dimensional vector with its $n$-th $\left(n \in \{(i-1)N+i|i\in \mathcal{N}\}\right)$ elements being one and the others being zero. Then, we find the desirable solution of $\mathbf{v}$ that minimizes $\mathrm{tr}\left(\left(\mathbf{C}^{T}\otimes \mathbf{B} \right) \mathbf{V}\right)$ among all random generated $\mathbf{v}$'s. Finally, let $\hat{v}_{n}$ denote the $n$-th element in $\mathbf{v}^{\star}$ and a efficient solution to problem (P4.2) is obtained as $\mathbf{\Phi}^{\star} = \mathrm{diag}\left(\hat{v}_{1},\ldots,\hat{v}_{(i-1)N+i},\ldots,\hat{v}_{N^{2}}\right)$.

\subsubsection{Optimization of $\mathbf{P}$}
Based on problem (P4.1), the optimization of $\mathbf{P}$ is reformulated as 
	\begin{eqnarray}
			&\!\!\!(\text{P4.4}):\!\!\!&  \nonumber\\
			&\!\!\!\!\!\!{\mathop {\min}\limits_{\mathbf{P} } }\!\!\!\!\!\!& \mathrm{tr}\left(\left({\mathbf{G}} \mathbf{R}_{\mathbf{x}} {\mathbf{G}^H} \right)^{\!\!-1} \!{\mathbf{P}}^{-2} \right) \mathrm{tr}\left( \left( {\mathbf{G}}{\bf{R}}_{\bf{w}}^{ - 1} {\mathbf{G}^H} \right)^{\!\!-1} \!{\mathbf{P}}^{-2} \right) \nonumber\\
		&\!\!\!\!\!\!\text{s.t.}\!\!\!\!\!\! & \eqref{IRS_power_cons1_eq3}~\text{and}~ \eqref{IRS_power_cons5_eq}.\nonumber
	\end{eqnarray}
In the objective function of problem (P4.4), $\mathbf{R}_{{\mathbf{w}}}^{-1} = (\sigma_{\text{r}}^{2}  \mathbf{G}^{T}\mathbf{P}^{2}\mathbf{G}^{*}+\sigma_{\text{b}}^{2}\mathbf{I}_{{M}})^{-1}$ is the inverse of a quadratic function of matrix $\mathbf{P} $, which is highly non-convex and challenging to optimize. To address this issue, we propose to iteratively update the variables $\mathbf{P}$ and $\mathbf{R}_{{\mathbf{w}}}$, by treating the other to be given. By defining $\mathbf{Q}=\mathbf{P}^{2}$, $\mathbf{T}_{1} = \left({\mathbf{G}} \mathbf{R}_{\mathbf{x}} {\mathbf{G}^H} \right)^{\!\!-1}$, $\mathbf{T}_{2} =  \left( {\mathbf{G}}{\bf{R}}_{\bf{w}}^{ - 1} {\mathbf{G}^H} \right)^{-1}$, $\mathbf{q} = \mathrm{Diagvec}(\mathbf{Q})$, $\mathbf{t}_{1} = \mathrm{Diagvec}(\mathbf{T}_{1})$, and $\mathbf{t}_{2} = \mathrm{Diagvec}(\mathbf{T}_{2})$, problem (P4.4) is equivalent to 
\begin{subequations}
	\begin{eqnarray}
		(\text{P4.5}):	& {\mathop {\min}\limits_{\mathbf{Q} } }\!\!\!\!& (\mathbf{q}^{-1})^{T}\mathbf{t}_{1}\mathbf{t}_{2}^{T}\mathbf{q}^{-1} \nonumber\\
		& \text{s.t.}\!\!\!\!\!\! & \mathrm{tr}\left(\mathbf{Q}\mathbf{E}\mathbf{Q}^{\frac{1}{2}}\mathbf{\Phi}\mathbf{T}_{1}^{-1}\mathbf{\Phi}^{H}\mathbf{Q}^{\frac{1}{2}}\mathbf{E}^{H}\right) \nonumber\\
		&& + \sigma_{\text{r}}^2\mathrm{tr}\left(\mathbf{Q}\mathbf{E}\mathbf{Q}\mathbf{E}^{H}\right)\!\!+\! \mathrm{tr}\left( \mathbf{T}_{1}^{-1}\mathbf{Q} \right)  \nonumber \\
		&& + 2 \sigma_{\text{r}}^2 \mathrm{tr}\left(\mathbf{Q}\right) \leq P_{\text{s}}, \label{IRS_power_cons1_eq4}\\
		&&[\mathbf{Q}]_{i,j} \left\{ \begin{array}{ll}
			\leq {a_{\text{max}}^{2}},& \forall i=j ,\\
			 = 0, &\forall i \neq j .
		\end{array} \right.\label{IRS_power_cons5_eq2}
	\end{eqnarray}
\end{subequations}
Note that problem (P4.5) is still non-convex due to the non-convexity of both the objective function and the constraint in \eqref{IRS_power_cons1_eq4}. Particularly, the optimal solution to problem (P4.5) can be directly obtained when $[\mathbf{Q}]_{i,i} = {a_{\text{max}}^{2}}, \forall i \in \mathcal{N}$ satisfies the constraint in \eqref{IRS_power_cons1_eq4}, as presented in Proposition \ref{prop5}. 
\begin{prop}\label{prop5}
	If $[\mathbf{Q}]_{i,i} = {a_{\text{max}}^{2}}, \forall i \in \mathcal{N}$, satisfies the constraint in \eqref{IRS_power_cons1_eq4}, the optimal solution to problem (P4.5) is given by
	\begin{align}
		[\mathbf{Q}]_{i,i} = {a_{\text{max}}^{2}}, \forall i \in \mathcal{N}.
	\end{align}
\end{prop}
\begin{IEEEproof}
	Let $t_{1,i}$, $t_{2,i}$, and $q_{i}$ denote the $i$-th elements in $\mathbf{t}_{1}$, $\mathbf{t}_{2}$, and $\mathbf{Q}$, respectively. Then, the objective function in problem P(4.5) is equivalent to 
	\begin{align}
		\left(\sum\nolimits_{i = 1}^{N} t_{1,i}q_i\right)\left(\sum\nolimits_{i=1}^{N} t_{2,i}q_i\right). \label{Obj_P5.5_expand}
	\end{align}
Since $\mathbf{T}_{1}$ and $\mathbf{T}_{2}$ are positive semi-definite matrices, we have $t_{1,i} \geq 0$ and $t_{2,i} \geq 0$. Then the derivative of the function \eqref{Obj_P5.5_expand} with respect to $q_{i'}$ is $t_{1,i'}\left(\sum_{i=1}^{N} t_{2,i}q_i\right) + \left(\sum_{i = 1}^{N} t_{1,i}q_i\right)t_{2,i'}$ which is clearly greater than or equal to zero. Thus, the objective function is a  monotonically non-decreasing function with respect to $\{q_i\}$, i.e., the diagonal elements of $\mathbf{Q}$. As a result, the optimal solution to problem (P4.5) is $[\mathbf{Q}]_{i,i} = {a_{\text{max}}^{2}}, \forall i \in \mathcal{N}$.
\end{IEEEproof}

 When the inequation in \eqref{IRS_power_cons1_eq4} does not holds with $[\mathbf{Q}]_{i,i} = {a_{\text{max}}^{2}}, \forall i \in \mathcal{N}$, we transform problem (P4.5) into a convex form to obtain an efficient solution by resorting to SCA. Specifically, we approximate the objective function in problem (P4.5) at the local point $\mathbf{Q}^{(l)}$ (i.e., $\mathbf{q}^{(l)}$) as: 
\begin{align}
	((\mathbf{q}^{(l)})^{-1})^{T}\mathbf{t}_{1}\mathbf{t}_{2}^{T}(\mathbf{q}^{(l)})^{-1} +  \mathrm{tr}\left(\mathbf{d}_{1}^{T}(\mathbf{q}^{(l)})\left(\mathbf{q} \!-\! \mathbf{q}^{(l)}\right)\right), \label{Obj_sensing_eq}
\end{align}
where $\mathbf{d}_{1}^{T}(\mathbf{q}) \!=\!  -	(\mathbf{q}^{-1})^{T}\mathbf{t}_{1}\left(\mathbf{t}_{2}^{T}\! \circ \! \mathbf{q}^{-2}\right) \!-\! (\mathbf{q}^{-1})^{T}\mathbf{t}_{2}\left(\mathbf{t}_{1}^{T}\! \circ \! \mathbf{q}^{-2}\right)$.
At a local point $\mathbf{Q}^{(l)}$, we approximate the first term in the constraint \eqref{IRS_power_cons1_eq4} as 
{\small\begin{align}
	&\mathrm{tr}\left(\widehat{\mathbf{C}}(\mathbf{Q}^{(l)})^{\frac{1}{2}}\mathbf{E}^{H}\mathbf{Q}^{(l)}\mathbf{E}(\mathbf{Q}^{(l)})^{\frac{1}{2}}\right) \!+\! \mathrm{tr}\left(\mathbf{D}_{2}^{T}(\mathbf{Q}^{(l)})\left(\mathbf{Q} \!-\! \mathbf{Q}^{(l)}\right)\!\right) , \label{approx1_sensing}
\end{align}}where $\widehat{\mathbf{C}} = \mathbf{\Phi}\mathbf{G}\mathbf{R}_{\mathbf{x}}\mathbf{G}^{H}\mathbf{\Phi}^{H}$ and 
\begin{align}
	\mathbf{D}_{2}(\mathbf{Q}) &= \mathbf{E}^{*}\mathbf{Q}^{\frac{1}{2}}\widehat{\mathbf{C}}^{T}\mathbf{Q}^{\frac{1}{2}}\mathbf{E}^{T}+\frac{1}{2}\mathbf{Q}^{-\frac{1}{2}}\widehat{\mathbf{C}}^{T}\mathbf{Q}^{\frac{1}{2}}\mathbf{E}^{T}\mathbf{Q}\mathbf{E}^{*}\nonumber\\
	&+\frac{1}{2}\mathbf{Q}^{-\frac{1}{2}}\mathbf{E}^{T}\mathbf{Q}\mathbf{E}^{*}\mathbf{Q}^{\frac{1}{2}}\widehat{\mathbf{C}}^{T}. \label{D_4}
\end{align}
Similarly, the second term in constraint \eqref{IRS_power_cons1_eq4} is approximated as
\begin{align}
	\!\!\! \sigma_{\text{r}}^2\mathrm{tr}\left(\mathbf{Q}^{(l)}\mathbf{E}\mathbf{Q}^{(l)}\mathbf{E}^{H}\right) \!+\! \sigma_{\text{r}}^2\mathrm{tr}\left(\mathbf{D}_{3}^{T}(\mathbf{Q}^{(l)})\left(\mathbf{Q}\!-\!\mathbf{Q}^{(l)}\right)\right),\label{approx2_sensing}
\end{align}
where $\mathbf{D}_{3}(\mathbf{Q}) = \mathbf{E}^{*}\mathbf{Q}\mathbf{E}^{T}+\mathbf{E}^{T}\mathbf{Q}\mathbf{E}^{*}$.
Based on \eqref{approx1_sensing} and \eqref{approx2_sensing}, the constraint in \eqref{IRS_power_cons1_eq4} is approximated as the following convex constraint:
\begin{align}
	&\mathrm{tr}\left(\widehat{\mathbf{C}}(\mathbf{Q}^{(l)})^{\frac{1}{2}}\mathbf{E}^{H}\mathbf{Q}^{(l)}\mathbf{E}(\mathbf{Q}^{(l)})^{\frac{1}{2}}\right) \nonumber\\
	&+ \mathrm{tr}\left(\mathbf{D}_{2}^{T}(\mathbf{Q}^{(l)})\left(	\mathbf{Q}-\mathbf{Q}^{(l)}\right)\right)  \nonumber\\
	& +\sigma_{\text{r}}^2\mathrm{tr}\left(\mathbf{Q}^{(l)}\mathbf{E}\mathbf{Q}^{(l)}\mathbf{E}^{H}\right) + \sigma_{\text{r}}^2\mathrm{tr}\left(\mathbf{D}_{3}^{T}(\mathbf{Q}^{(l)})\left(\mathbf{Q}-\mathbf{Q}^{(l)}\right)\right)\nonumber\\
	&+\! \mathrm{tr}\left( \mathbf{T}_{1}^{-1}\mathbf{Q} \right)  + 2 \sigma_{\text{r}}^2 \mathrm{tr}\left(\mathbf{Q}\right) \leq P_{\text{s}}, \label{IRS_power_cons1_eq4_approx}
\end{align}

As a result, problem (P4.5) is approximated into the following problem at the $l$-th SCA iteration:
	\begin{eqnarray}
			&\!\!\!\!\!\!(\text{P4.5.$l$}):\!\!\!&  \nonumber\\
			&\!\!\!\!\!\!{\mathop {\min}\limits_{\mathbf{Q} } }\!\!\!\!\!\!&\!\!\!\!\! 	((\mathbf{q}^{(l)})^{-1})^{T}\mathbf{t}_{1}\mathbf{t}_{2}^{T}(\mathbf{q}^{(l)})^{-1} \!\!+\!  \mathrm{tr}\!\left(\mathbf{d}_{1}^{T}(\mathbf{q}^{(l)})\left(\mathbf{q} \!-\! \mathbf{q}^{(l)}\right)\right)  \nonumber\\
		&\!\!\!\!\!\!\text{s.t.}\!\!\!\!\!\! & \eqref{IRS_power_cons5_eq2}~\text{and}~\eqref{IRS_power_cons1_eq4_approx}. \nonumber
	\end{eqnarray}
Problem (P4.5.$l$) is a convex problem that can be efficiently solved using CVX. Let $\mathbf{Q}^{\star(l)}$ denote the obtained solution to problem (P4.5.$l$) at the $l$-th iteration, which is then updated as the local point $\mathbf{Q}^{(l+1)}$ for the $(l+1)$-th iteration. It is important to note that the objective value achieved by $\mathbf{Q}^{\star(l+1)}$ is always no greater than that achieved by $\mathbf{Q}^{\star(i)}$, indicating that the CRB is monotonically non-increasing after each iteration of the SCA. Furthermore, the optimal value of problem (P4.5) is lower-bounded due to the non-negativity of the CRB value. Thus, the convergence of the SCA is guaranteed, and we denote $\mathbf{Q}^{\star}$ as the corresponding converged solution. Consequently, we obtain the optimal solution to (P4.4) as $\mathbf{P}^{\star} = ({\mathbf{Q}^{\star}})^{-\frac{1}{2}}$. By alternately optimizing $\bm{\Phi}$ and $\mathbf{P}$ based on (P4.2) and (P4.4) until convergence, we obtain an efficient solution $\mathbf{\Psi}^{\star}$ to problem (P4).

In summary, the AO-based algorithm for solving problem (P2) is implemented by alternately solving problems (P3) and (P4). Since problem (P3) is optimally solved and the solution to problem (P4) yields a decreasing CRB, the algorithm ensures that the CRB value is monotonically non-increasing over iterations, thereby guaranteeing convergence.

\section{Joint Beamforming Design for ISAC scenario}
In this section, we consider the general ISAC scenario and jointly design the transmit beamforming $\{\mathbf{R}_{0},\mathbf{w}_k\}$ at the BS and the reflective beamforming $\mathbf{\Psi}$ at the IRS to enhance the performance of target estimation. Specifically, we propose an efficient algorithm to solve problem (P1). 
\subsection{Optimal Information Signal beamforming and Sensing Signal Convariance with Given $\mathbf{\Psi}$}
Under given reflective beamforming $\mathbf{\Psi}$, the optimization of transmit beamforming is reformulated as 
\begin{align}
	\begin{array}{*{20}{lcl}}
		\!\!(\text{P5}):&\!\!\!\!\!{\mathop {\min}\limits_{\{\mathbf{w}_{k},\mathbf{R}_{0}\} } }\!\!\!\!\!\! &\mathrm{tr}\left(\left({\mathbf{G}} \left(\sum\nolimits_{k\in \mathcal{K}}\mathbf{w}_{k}\mathbf{w}_{k}^{H} + \mathbf{R}_{0}\right) {\mathbf{G}^H} \right)^{\!\!-1}\!{\mathbf{P}}^{-2} \right)  \nonumber\\
		&\!\!\!\!\!\text{s.t.} \!\!\!\!\!\! & \eqref{P1_I_cons1}-\eqref{P1_I_cons3}.\nonumber
\end{array}
\end{align}
By defining  $ {\mathbf{H}}_{k} =  {\mathbf{h}}_{k} {\mathbf{h}}_{k}^{H}$, $\bar{\mathbf{H}}_{k} = \bar{\mathbf{h}}_{k}\bar{\mathbf{h}}_{k}^{H}$, and $\mathbf{W}_{k} = \mathbf{w}_{k}\mathbf{w}_{k}^{H}$ with $\mathrm{rank}\left(\mathbf{W}_{k}\right) =1$, the constraint in \eqref{P1_I_cons1} is equivalent to 
\begin{align}
	&\sum\nolimits_{k' \in \mathcal{K} \backslash \{k\}} \mathrm{tr}\left(\bar{\mathbf{H}}_{k}^{H}\mathbf{W}_{k'}\right) -\frac{1}{\Gamma_{k}} \mathrm{tr}\left(\bar{\mathbf{H}}_{k}^{H}\mathbf{W}_{k}\right) + \mathrm{tr}\left(\bar{\mathbf{H}}_{k}^{H} \mathbf{R}_{0}\right) \nonumber\\
	&+\sigma_{\text{r}}^2\mathrm{tr}\left(\mathbf{H}_{k}\mathbf{P}^{2}\right)  \!+ \! \sigma_{\text{u}}^2 \leq 0, \forall k \in \mathcal{K}. \label{P1_I_cons1_eq}
\end{align}
Then, by dropping the rank-one constraint, problem (P5) is equivalent to 
	\begin{eqnarray}
			&\!\!\!(\text{SDR5.1}):\!\!\!&  \nonumber\\
			&\!\!\!\!\!\!{\mathop {\min}\limits_{\{\mathbf{W}_{k},\mathbf{R}_{0}\}} }\!\!\!\!\!\!& \mathrm{tr}\left(\left({\mathbf{G}} \left(\sum\nolimits_{k\in \mathcal{K}}\mathbf{W}_{k} + \mathbf{R}_{0}\right) {\mathbf{G}^H} \right)^{\!\!-1}\!{\mathbf{P}}^{-2} \right) \nonumber\\
		&\!\!\!\!\!\!\text{s.t.}\!\!\!\!\!\!\!\! &\!\!\!\!\!\!  \eqref{P1_I_cons2}-\eqref{P1_I_cons3}, ~\text{and}~\eqref{P1_I_cons1_eq}.\nonumber
	\end{eqnarray}
Note that problem (SDR5.1) is a convex problem that can be solved by CVX.  Let $\{\widetilde{\mathbf{W}}_{k}\}$ and $\widetilde{\mathbf{R}_{0}}$ denote the obtained optimal solution to problem (SDR5.1). The optimal rank-one solution $\{\widetilde{\mathbf{W}}_{k}^{\star}\}$ and the corresponding $\widetilde{\mathbf{R}}_{0}^{\star}$ to problems (P5) and (P5.1) can be constructed via the following proposition.
\begin{prop}\label{prop4}
	Based on the obtained optimal solution $\{\widetilde{\mathbf{W}}_{k}\}$ and $\widetilde{\mathbf{R}_{0}}$ to (SDR5.1), the optimal  solution of $\{\mathbf{W}_{k}\}$ to problems (P5) and (P5.1) is given by 
	\begin{align}
		\widetilde{\mathbf{W}}_{k}^{\star} = \widetilde{\mathbf{w}}_{k}\widetilde{\mathbf{w}}_{k}^{H}, \forall k \in \mathcal{K}, \label{prop4.1}
	\end{align}
	with $\widetilde{\mathbf{w}}_{k} = \left(\bar{\mathbf{h}}_{k}^{H}\widetilde{\mathbf{W}}_{k}\bar{\mathbf{h}}_{k}\right)^{-\frac{1}{2}}\widetilde{\mathbf{W}}_{k}\bar{\mathbf{h}}_{k}$, and the corresponding ${\mathbf{R}}_{0}^{\star}$ is given by 
	\begin{align}		
		\widetilde{\mathbf{R}}_{0}^{\star} = \widetilde{\mathbf{R}}_{0}+ \sum\nolimits_{k \in \mathcal{K}}\widetilde{\mathbf{W}}_{k}-\sum\nolimits_{k \in \mathcal{K}}\widetilde{\mathbf{W}}_{k}^{\star}.\label{prop4.2}
	\end{align}
\end{prop}
\begin{IEEEproof}
\textcolor{black}{It is observed  that the reconstructed solution does not change the value of sample covariance
matrix $\mathbf{R}_{\mathbf{x}}$. Thus, the alternative solution satisfies constraints \eqref{P1_I_cons2} and \eqref{P1_I_cons3}. Meanwhile, it can be proved that the reconstructed beamformers $\widetilde{\mathbf{W}}_{k}^{\star}$ and $\widetilde{\mathbf{R}}_{0}^{\star}$ are semi-definite and satisfy the SINR constraint in \eqref{P1_I_cons1_eq}.} The detailed proof is similar to \cite[Proposition 5]{fang2023multiirsenabled} and thus omitted for brevity.
\end{IEEEproof}

It is shown in Proposition \ref{prop4} that the solutions $\{\widetilde{\mathbf{W}}_{k}^{\star}\}$ and $\widetilde{\mathbf{R}}_0^{\star}$ in \eqref{prop4.1} and \eqref{prop4.2} are actually optimal for problem (SDR5.1). Therefore, the SDR is tight between problems (P5) and (SDR5.1), and thus we obtain the optimal solution to problem (P5). 
\subsection{Reflective Beamforming Optimization with Given $\{\mathbf{w}_{k},\mathbf{R}_{0}\}$}
Under given transmit beamforming $\{\mathbf{w}_{k},\mathbf{R}_{0}\}$, the optimization of reflective beamforming is reformulated as
\begin{align}
\begin{array}{*{20}{lcl}}
	(\text{P6}):&{\mathop {\min}\limits_{\mathbf{\Psi}} }\!\!\!\! &\mathrm{tr}\left(\mathbf{T}_{1}{\mathbf{P}}^{-2} \right) \mathrm{tr}\left( \mathbf{T}_{2} {\mathbf{P}}^{-2} \right) \nonumber\\
	&\text{s.t.}\!\!\!\! & \eqref{P1_I_cons1},\eqref{P1_I_cons2}, ~\text{and}~\eqref{P1_I_cons5}.\nonumber
\end{array}
\end{align}
Note that the objective function in problem (P6) is only dependent on $\mathbf{P}$, thus we propose to optimize the magnitude and phase of the reflection coefficients separately.
\subsubsection{Optimization of $\mathbf{\Phi}$}
First, we optimize $\mathbf{\Phi}$ with a given $\mathbf{P}$. As the objective function in problem (P6) is independent of $\mathbf{\Phi}$, we propose to optimize $\mathbf{\Phi}$ to maximize the minimum SINR among all CUs. Based on the definition $\mathbf{\Psi } = \mathbf{P}\mathbf{\Phi}$, problem (P6) is reformulated into
	\begin{eqnarray}
		&\!\!\!\!\!\!\!(\text{P6.1}):\!\!\!\!\!\!\!\!&\nonumber\\
		&\!\!\!\!\!\!\!\!\!{\mathop {\max}\limits_{\mathbf{\Phi}} }\!\!\!\! & {\mathop {\min} \limits_{k\in \mathcal{K}}} \; \frac{ \left|\bar{\mathbf{h}}_{k}^{H}\mathbf{w}_{k}\right|^2 }{{\sum\limits_{k' \in \mathcal{K} \backslash \{k\}} \left|\bar{\mathbf{h}}_{k}^{H}\mathbf{w}_{k'}\right|^2  \!+\! \bar{\mathbf{h}}_{k}^{H} \mathbf{R}_{0}\bar{\mathbf{h}}_{k}  \!+\!  \sigma_{\text{r}}^2\mathbf{h}_{k}^{H} \mathbf{P}^{2}\mathbf{h}_{k} \!+\! \sigma_{\text{u}}^2}} \nonumber\\
		&\!\!\!\!\!\!\!\!\!\text{s.t.}\!\!\!\! &   \eqref{IRS_power_cons1_eq3}, \eqref{Phi_mod_cons}, ~\text{and}~\eqref{Phi_semi_cons}. \nonumber
\end{eqnarray}
Problem (P6.1) is a non-convex problem due to the non-convexity of the objective function and the constraint in \eqref{IRS_power_cons1_eq3}. To solve this problem, we first introduce an auxiliary variable $\kappa$ to transform problem (P6.1) into the following problem:
{\begin{subequations}
 	\begin{eqnarray}
 		&\!\!\!\!\!\!\!\!\!\!\!\!(\text{P6.2}):\!\!\!\!\!&\nonumber\\
 		&\!\!\!\!\!\!\!\!\!\!\!\!\!\!\!\!\!\!{\mathop {\max}\limits_{\mathbf{\Phi},\kappa} }\!\!\!\!\!\!\! & \kappa \nonumber\\
 		&\!\!\!\!\!\!\!\!\!\!\!\!\!\!\!\!\!\!\text{s.t.}\!\!\!\!\!\!\! &\!\!\!   \frac{ \left|\bar{\mathbf{h}}_{k}^{H}\mathbf{w}_{k}\right|^2 }{{\sum\nolimits_{k' \in \mathcal{K} \backslash \{k\}}\!\! \left|\bar{\mathbf{h}}_{k}^{H}\mathbf{w}_{k'}\right|^2 \!\!\! +\! \bar{\mathbf{h}}_{k}^{H} \mathbf{R}_{0}\bar{\mathbf{h}}_{k}  \!+\!  \sigma_{\text{r}}^2\mathbf{h}_{k}^{H} \mathbf{P}^{2}\mathbf{h}_{k} \!+\! \sigma_{\text{u}}^2}}\! \nonumber\\
		&&\geq \kappa, \label{sinr_cons_eq} \\
 		&&\eqref{IRS_power_cons1_eq3}, \eqref{Phi_mod_cons},~\text{and}~\eqref{Phi_semi_cons}. \nonumber
 \end{eqnarray}
 \end{subequations}}Then, by defining  $\breve{\mathbf{H}}_{k} = \mathrm{diag}\left(\mathbf{h}_{k}^{H}\right)$ and $\mathbf{\Theta} = \bm{\phi}^{*}\bm{\phi}^{T}$, the constraints in \eqref{IRS_power_cons1_eq3} and \eqref{sinr_cons_eq} are transformed into 
\begin{align}
	& 	\mathrm{tr}\left(\left(\mathbf{C}^{T}\otimes \mathbf{B} \right) \mathbf{V}\right)  + \sigma_{\text{r}}^2\mathrm{tr}\left(\mathbf{P}^{2}\mathbf{E}\mathbf{P}^{2}\mathbf{E}^{H}\right)\!\!+\! \mathrm{tr}\left( \mathbf{P}^{2}\mathbf{G}\mathbf{R}_{\mathbf{x}}\mathbf{G}^{H} \right)  \nonumber \\
	& + 2 \sigma_{\text{r}}^2 \mathrm{tr}\left(\mathbf{P}^{2}\right) \leq P_{\text{s}},\label{IRS_power_cons1_eq5}
\end{align}
and
{\small
	\begin{align}
		&\sum\limits_{k' \in \mathcal{K} \backslash \{k\}}\!\!\!\mathrm{tr} \left( \breve{\mathbf{H}}_{k}^{H}\mathbf{G}\mathbf{W}_{k'}\mathbf{G}^{H}\breve{\mathbf{H}}_{k}\mathbf{P}\mathbf{\Theta}\mathbf{P}\right) \!+\! \mathrm{tr} \left(\breve{\mathbf{H}}_{k}^{H}\mathbf{G} \mathbf{R}_{0}\mathbf{G}^{H}\breve{\mathbf{H}}_{k}\mathbf{P}\mathbf{\Theta}\mathbf{P}\right)  \nonumber\\
		&\!-\!\frac{1}{\kappa}\mathrm{tr} \left( \breve{\mathbf{H}}_{k}^{H}\mathbf{G}\mathbf{W}_{k}\mathbf{G}^{H}\breve{\mathbf{H}}_{k}\mathbf{P}\mathbf{\Theta}\mathbf{P}\right)\!+\! \sigma_{\text{r}}^2\mathrm{tr} \left(\mathbf{H}_{k}\mathbf{P}^{2}\right) \!+\! \sigma_{\text{u}}^2 \! \leq \! 0, \forall k \in \mathcal{K},\label{sinr_cons_eq2}
\end{align}}respectively. Recall the definition of $\mathbf{V}$ in Section IV-B, the relationship between $\mathbf{V}$ and $\mathbf{\Theta}$ is given by
\begin{align}
	[\mathbf{V}]_{(i-1)N+i,(j-1)N+j} =  [\mathbf{\Theta}]_{i,j}, \forall i,j\in \mathcal{N}, \label{V_Theta}
\end{align}
 and the other elements in $\mathbf{V}$ are equal to zero. As such, by using the SDR technique and dropping the rank-one constraint, problem (P6.1) is transformed into the following problem:
		\begin{eqnarray}
			(\text{SDR6.2}):&{\mathop {\max}\limits_{\mathbf{V},\kappa} }& \kappa \nonumber\\
			&\text{s.t.} & \eqref{V_mod_cons}, \eqref{V_semi_cons},~\text{and}~\eqref{IRS_power_cons1_eq5}-\eqref{V_Theta}. \nonumber
	\end{eqnarray}
Let $\mathbf{V}^{\divideontimes}$ denote the optimal solution to problem (SDR6.2) and then we employ a bisection method to obtain it. Specifically, we first define the search range of the objective value to problem (\text{SDR6.2}) as $\kappa \in [\kappa_{\text{min}},\kappa_{\text{max}}]$ and a tolerance $\epsilon$. In each iteration $\varsigma$, let $\kappa^{(\varsigma)} = \frac{\kappa_{\text{min}}+\kappa_{\text{max}}}{2}$ and then we need to check the following feasibility problem: 
		\begin{eqnarray}
			(\text{SDR6.3}):&\text{Find}& \mathbf{V}\nonumber\\
			&\text{s.t.} & \eqref{V_mod_cons}, \eqref{V_semi_cons},~\text{and}~\eqref{IRS_power_cons1_eq5}-\eqref{V_Theta}. \nonumber
	\end{eqnarray}
If problem (SDR6.3) is feasible, we denote $\mathbf{V}^{(\varsigma)}$ as the solution to the problem and accordingly set $\kappa_{\text{min}} = {\kappa}^{(\varsigma)}$ and $\mathbf{V}^{\divideontimes}=\mathbf{V}^{(\varsigma)}$. Otherwise, we set $\kappa_{\text{max}} = {\kappa}^{(\varsigma)}$. We perform the above process iteratively until the condition $\kappa_{\text{max}}-\kappa_{\text{min}}\leq \epsilon$ is satisfied. Finally, we obtain the optimal solution $\mathbf{V}^{\divideontimes}$ to problem (SDR6.2) and recover the optimal $\mathbf{\Theta}^{\star}$ according to \eqref{V_Theta}. Next, we need to find a high-quality solution to problem (P6.2). Since the obtained $\mathbf{\Theta}^{\star}$ may not be rank-one, it might not be the optimal solution to problem (P6.1). To address this, we use the Gaussian randomization method to construct a high-quality rank-one solution for problem (P6.1).  We first generate a number of random vectors ${\mathbf{r}} \sim \mathcal{CN}\left(\mathbf{0},\mathbf{I}_N\right)$, and construct a number of rank-one candidates as $\mathbf{\phi} = e^{\jmath\mathrm{arg}\left\{\left(\mathbf{\Theta}^{\star}\right)^{\frac{1}{2}}\mathbf{r}\right\}}$. 
We then find the desired $\mathbf{\phi}$ that maximizes the objective function in problem (P6.1) among all randomly generated $\mathbf{\phi}$'s.  As a result, problem (P6.1) is finally solved. 
\subsubsection{Optimization of $\mathbf{P}$}
Then, we optimize  $\mathbf{P}$ under a given $\mathbf{\Phi}$. In this case, the optimization problem is formulated as
	\begin{eqnarray}
		(\text{P6.4}):	&{\mathop {\min}\limits_{\mathbf{P} } }& \mathrm{tr}\left(\mathbf{T}_{1} {\mathbf{P}}^{-2} \right) \mathrm{tr}\left( \mathbf{T}_{2} {\mathbf{P}}^{-2} \right) \nonumber\\
		&\!\!\!\!\!\!\text{s.t.}\!\!\!\!\!\! &  \eqref{IRS_power_cons1_eq3}, \eqref{IRS_power_cons5_eq},~\text{and}~\eqref{sinr_cons_eq}.\nonumber
	\end{eqnarray}
Similar to problem (P4.4), we first isolate $\mathbf{R}_{{\mathbf{w}}}$ from the objective function, and then iteratively update the variables $\mathbf{P}$ and $\mathbf{R}_{{\mathbf{w}}}$ until convergence is achieved. By defining $\bar{\mathbf{P}}=\mathbf{p}\mathbf{p}^{T}$, we have $\mathrm{rank}(\bar{\mathbf{P}}) = 1$ and $[\bar{\mathbf{P}}]_{i,i}\leq a_{\text{max}}^{2}$. Then, the constraint in \eqref{sinr_cons_eq} is equivalent to 
\begin{align}
		&\sum\nolimits_{k' \in \mathcal{K} \backslash \{k\}}\!\!\!\!\mathrm{tr}\! \left( \mathbf{\Phi}\breve{\mathbf{H}}_{k}^{H}\mathbf{G}\mathbf{W}_{k'}\mathbf{G}^{H}\breve{\mathbf{H}}_{k}\mathbf{\Phi}^{H}\bar{\mathbf{P}}\right) \nonumber\\
		&+\mathrm{tr}\! \left(\mathbf{\Phi}\breve{\mathbf{H}}_{k}^{H}\mathbf{G} \mathbf{R}_{0}\mathbf{G}^{H}\breve{\mathbf{H}}_{k}\mathbf{\Phi}^{H}\bar{\mathbf{P}}\right) \!+\! \sigma_{\text{r}}^2\mathrm{tr} \left(\mathbf{H}_{k}\bar{\mathbf{P}}\right) \!+\! \sigma_{\text{u}}^2 \! \nonumber\\
		&-\frac{1}{\Gamma_{k}}\mathrm{tr}\! \left(\mathbf{\Phi} \breve{\mathbf{H}}_{k}^{H}\mathbf{G}\mathbf{W}_{k}\mathbf{G}^{H}\breve{\mathbf{H}}_{k}\mathbf{\Phi}^{H}\bar{\mathbf{P}}\right) \leq \! 0, \forall k \in \mathcal{K}.\label{sinr_cons_eq4}
\end{align}
We define $\bar{\mathbf{p}} = \mathrm{Diagvec}(\bar{\mathbf{P}})$ and then the objective function in problem (P6.4) is reformulated as $(\bar{\mathbf{p}}^{-1})^{T}\mathbf{t}_{1}\mathbf{t}_{2}^{T}\bar{\mathbf{p}}^{-1}$. 
Let $\bar{\mathbf{P}}^{(l)}$ denote the local point for Taylor approximation at the $l$-th SCA iteration, the objective function in problem (P6.4) is approximated as $((\bar{\mathbf{p}}^{(l)})^{-1})^{T}\mathbf{t}_{1}\mathbf{t}_{2}^{T}(\bar{\mathbf{p}}^{(l)})^{-1} +  \mathrm{tr}\left(\mathbf{d}_{2}^{T}(\bar{\mathbf{p}}^{(l)})\left(\bar{\mathbf{p}} \!-\! \bar{\mathbf{p}}^{(l)}\right)\right)$.
By defining $\bar{\mathbf{P}}_{d} = \mathrm{Diag}(\bar{\mathbf{P}})$, the constraint in \eqref{IRS_power_cons1_eq3} is equivalent to
\begin{align}
	&\mathrm{tr}\left(\bar{\mathbf{P}}_{d}\mathbf{E}\bar{\mathbf{P}}_{d}^{\frac{1}{2}}\widehat{\mathbf{C}}\bar{\mathbf{P}}_{d}^{\frac{1}{2}}\mathbf{E}^{H}\right)  + \sigma_{\text{r}}^2\mathrm{tr}\left(\bar{\mathbf{P}}_{d}\mathbf{E}\bar{\mathbf{P}}_{d}\mathbf{E}^{H}\right)\nonumber\\
	&+ \mathrm{tr}\left( \mathbf{T}_{1}^{-1}\bar{\mathbf{P}}_{d} \right)  + 2 \sigma_{\text{r}}^2 \mathrm{tr}\left(\bar{\mathbf{P}}_{d}\right) \leq P_{\text{s}}. \label{IRS_power_cons1_eq6}
\end{align}
Similar to \eqref{IRS_power_cons1_eq4_approx}, the constraint in \eqref{IRS_power_cons1_eq6} is approximated as:
\begin{align}
		&\mathrm{tr}\left(\widehat{\mathbf{C}}(\bar{\mathbf{P}}_{d}^{(l)})^{\frac{1}{2}}\mathbf{E}^{H}(\bar{\mathbf{P}}_{d}^{(l)})\mathbf{E}(\bar{\mathbf{P}}_{d}^{(l)})^{\frac{1}{2}}\right)+ 2 \sigma_{\text{r}}^2 \mathrm{tr}\left(\bar{\mathbf{P}}_{d}\right) \nonumber\\
	&+\! \mathrm{tr}\left(\mathbf{D}_{2}^{T}(\bar{\mathbf{P}}_{d}^{(l)})\left(\bar{\mathbf{P}}_{d} \!-\! \bar{\mathbf{P}}_{d}^{(l)}\right)\right) + \sigma_{\text{r}}^2\mathrm{tr}\left(\bar{\mathbf{P}}_{d}^{(l)}\mathbf{E}\bar{\mathbf{P}}_{d}^{(l)}\mathbf{E}^{H}\right)\nonumber\\
	& +\sigma_{\text{r}}^2\mathrm{tr}\left(\mathbf{D}_{5}^{T}(\bar{\mathbf{P}}_{d})\left(\bar{\mathbf{P}}_{d}-\bar{\mathbf{P}}_{d}^{(l)}\right)\right) + \mathrm{tr}\left( \mathbf{T}_{1}^{-1}\bar{\mathbf{P}}_{d} \right) \leq P_{\text{s}}. \label{IRS_power_eq_isac} 
\end{align}
 As such, problem (P6.4) is reformulated as problem (SDR6.4) in the $l$-th SCA iteration by dropping the rank-one constraint of $\bar{\mathbf{P}}$.
\begin{subequations}
	\begin{eqnarray}
			&\!\!\!\!\!\!\!\!(\text{SDR6.4.$l$}):\!\!\!\!\!\!\!\!&  \nonumber\\
			&\!\!\!\!\!\!\!\!\!\!\!\!\!\!\!\!\!\!\!\!\!\!\!\!{\mathop {\min}\limits_{\bar{\mathbf{P}} } }\!\!\!\!\!\!\!\!&\!\!\!\!\!\!\!\!\!\!\!\!	((\bar{\mathbf{p}}^{(l)})^{-1})^{T}\mathbf{t}_{1}\mathbf{t}_{2}^{T}(\bar{\mathbf{p}}^{(l)})^{-1} +  \mathrm{tr}\left(\mathbf{d}_{2}^{T}(\bar{\mathbf{p}}^{(l)})\left(\bar{\mathbf{p}} \!-\! \bar{\mathbf{p}}^{(l)}\right)\right)  \nonumber\\
		&\!\!\!\!\!\!\!\!\!\!\!\!\!\!\!\!\!\!\!\!\!\!\!\!\text{s.t.}\!\!\!\!\!\!\!\!\!\!&\!\!\!\!\!\!\!\!\!\!\!\! \eqref{sinr_cons_eq4}~\text{and}~\eqref{IRS_power_eq_isac},\nonumber\\
		&&\!\!\!\!\!\!\!\!\!\!\!\!\bar{\mathbf{P}} \geq 0,\\
		&&\!\!\!\!\!\!\!\!\!\!\!\![\bar{\mathbf{P}}]_{i,i}\leq a_{\text{max}}^{2}.
	\end{eqnarray}
\end{subequations}
Notice that problem (SDR6.4.$l$) is a convex problem that can be solved by CVX. Let $\bar{\mathbf{P}}^{\divideontimes (l)}$ denote the optimal solution to problem (SDR6.4.$l$) in the $l$-th SCA iteration, which is updated as local point $\bar{\mathbf{P}}^{(l+1)}$ in the $(l+1)$-th SCA iteration. Let $\bar{\mathbf{P}}^{\divideontimes}$ denote the converged solution. Since the obtained $\bar{\mathbf{P}}^{\divideontimes}$ may not be rank-one, it might not be the optimal solution to problem (P6.4). 

Next, we use the Gaussian randomization method to construct a high-quality rank-one solution for problem (P6.4). The eigenvalue decomposition (EVD) of $\bar{\mathbf{P}}^{\divideontimes}$ is expressed as $\bar{\mathbf{P}}^{\divideontimes} = \mathbf{U}_{3}\mathbf{\Sigma}_{p}\mathbf{U}_{3}^{T}$. Then we generate a number of random vectors $\dot{\mathbf{r}} \sim \mathcal{N}\left(\mathbf{0},\mathbf{I}_N\right)$, and construct a number of rank-one candidates as $\dot{\mathbf{p}} = \mathrm{abs}(\mathbf{U}_{3}\mathbf{\Sigma}_{p}^{\frac{1}{2}}\dot{\mathbf{r}})$, where $\mathrm{abs}(\cdot)$ denotes the absolute value operator. However, the constructed $\dot{\mathbf{p}}$ may not be feasible for problem (P6.4). To address this issue, we introduce a scaling factor ${\tau_{p}}>0$ and find a feasible amplification coefficients vector $\sqrt{\tau_{p}}\dot{\mathbf{p}}$ for problem (P6.4).
By defining $\dot{\mathbf{P}} = \mathrm{diag}(\dot{\mathbf{p}})$, ${\ddot{\mathbf{P}}} =\dot{\mathbf{p}}\dot{\mathbf{p}}^{T}$, and substituting $\mathbf{p} = \sqrt{\tau_{p}}\dot{\mathbf{p}}$ into problem (P6.4), we can obtain a suitable ${\tau_{p}}$ by solving  the following optimization problem:
 \begin{subequations}
	\begin{eqnarray}
			\!\!\!\!\!\!\!	(\text{P6.5}):	& \!\!\!\!{\mathop {\min}\limits_{ \tau_{p} >0 } }\!\!\!\!&  \tau_{p}^{-2} \mathrm{tr}\left(\mathbf{T}_{1} \dot{\mathbf{P}}^{-2} \right) \mathrm{tr}\left( \mathbf{T}_{2} \dot{\mathbf{P}}^{-2} \right)  \nonumber\\
		& \!\!\!\!\text{s.t.}\!\!\!\!\!\! &\beta_{1}\tau_{p}^{2}  + \beta_{2}\tau_{p}^{2}  +  \beta_{3}\tau_{p}   + \beta_{4}\tau_{p} \leq P_{\text{s}}, \label{P7_6_cons1}\\
		&& \tau_{p}\max \left\{\mathrm{diag}\left(\dot{\mathbf{P}}^{2}\right)\right\} \leq a_{\text{max}}^{2},\label{P7_6_cons2}\\
		&&\tau_{p}\lambda_{1,k} + \sigma_{\text{u}}^2 \leq 0, \forall k \in \mathcal{K}, \label{P7_6_cons3}
	\end{eqnarray}
\end{subequations}
where $\beta_{1} = \mathrm{tr}\left(\dot{\mathbf{P}}^{2}\mathbf{E}\dot{\mathbf{P}}\widehat{\mathbf{C}}\dot{\mathbf{P}}\mathbf{E}^{H}\right)$, $\beta_{2} = \sigma_{\text{r}}^2\mathrm{tr}\left(\dot{\mathbf{P}}^{2}\mathbf{E}\dot{\mathbf{P}}^{2}\mathbf{E}^{H}\right)$, $\beta_{3} = \mathrm{tr}\left( \mathbf{T}_{1}^{-1}\dot{\mathbf{P}}^{2} \right)$, $ \beta_{4} = 2\sigma_{\text{r}}^2 \mathrm{tr}\left(\dot{\mathbf{P}}^{2}\right)$, and 
\begin{align}
	\lambda_{1,k} = &\sum\nolimits_{k' \in \mathcal{K} \backslash \{k\}}\!\!\!\!\mathrm{tr}\! \left( \mathbf{\Phi}\breve{\mathbf{H}}_{k}^{H}\mathbf{G}\mathbf{W}_{k'}\mathbf{G}^{H}\breve{\mathbf{H}}_{k}\mathbf{\Phi}^{H}{\ddot{\mathbf{P}}}\right) \nonumber\\
	&+\mathrm{tr}\! \left(\mathbf{\Phi}\breve{\mathbf{H}}_{k}^{H}\mathbf{G} \mathbf{R}_{0}\mathbf{G}^{H}\breve{\mathbf{H}}_{k}\mathbf{\Phi}^{H}{\ddot{\mathbf{P}}}\right) \!+\! \sigma_{\text{r}}^2\mathrm{tr} \left(\mathbf{H}_{k}{\ddot{\mathbf{P}}}\right)  \nonumber\\
	&-\frac{1}{\Gamma_{k}}\mathrm{tr}\! \left(\mathbf{\Phi} \breve{\mathbf{H}}_{k}^{H}\mathbf{G}\mathbf{W}_{k}\mathbf{G}^{H}\breve{\mathbf{H}}_{k}\mathbf{\Phi}^{H}{\ddot{\mathbf{P}}}\right).
\end{align}
Problem (P6.5) is equivalent to the following convex problem:
\vspace{-1mm}
 \begin{subequations}
	\begin{eqnarray}
			\!\!\!\!\!\!\!	(\text{P6.6}):	& \!\!\!\!{\mathop {\max}\limits_{ \tau_{p} >0 } }\!\!\!\!&  \tau_{p}^{2}   \nonumber\\
		& \!\!\!\!\text{s.t.}\!\!\!\!\!\! &(\beta_{1}+ \beta_{2})\tau_{p}^{2}  +  (\beta_{3} + \beta_{4})\tau_{p} \leq P_{\text{s}}, \label{P7_7_cons1}\\
		&& \tau_{p} \leq \frac{a_{\text{max}}^{2}}{\max \{\dot{\mathbf{p}}^{2}\}},\label{P7_7_cons2}\\
		&&\tau_{p} \geq -\frac{\sigma_{\text{u}}^2}{\lambda_{1,k}}, \forall k \in \mathcal{K}, \label{P7_7_cons3}
	\end{eqnarray}
\end{subequations}
It is observed that the objective function in problem (P6.6) is monotonically increasing with respect to the optimization variable $\tau_{p}$. Hence, the optimal  $\tau_{p}$ is located at the margin of the feasible region and given by $\tau_{p}^{\star} = \min \left\{\Lambda_{1},\frac{a_{\text{max}}^{2}}{\max \{\dot{\mathbf{p}}^{2}\}}\right\}$, where $\Lambda_{1} = \frac{\sqrt{\beta_{2}^{2} + 2\beta_{2}\beta_{4}+\beta_{4}^{2} +4\beta_{1}P_{\text{s}}  + 4\beta_{2}P_{\text{s}} }- \beta_{3} - \beta_{4} }{2\beta_{1} + 2\beta_{2} }$. As a result, a feasible solution to problem (P6.4) is given by $\sqrt{\tau_{p}^{\star}}\dot{\mathbf{p}}$. It should be noted that problem (P6.5) may not always be feasible. Consequently, it is necessary to perform multiple iterations of Gaussian randomization and select a feasible solution that minimizes the resulting CRB.
By alternately optimizing $\bm{\Phi}$ and $\mathbf{P}$ based on problems (P6.1) and (P6.4) until convergence, we obtain an effective solution $\mathbf{\Psi}^{\divideontimes}$ to problem (P6).

In summary, the AO-based algorithm for solving problem (P1) is implemented by alternately addressing problems (P5) and (P6). As problem (P5) is optimally solved and the solution to problem (P6) leads to a decreasing CRB, the algorithm ensures monotonically non-increasing CRB values over iterations, thus guaranteeing convergence.
\vspace{-2mm}
\section{Scaling Law Analysis}
In this section, we explore the inherent relationship between the power scaling of $\mathbf{R}_{\mathbf{x}}$ and the amplification scaling of $\mathbf{P}$ to provide more insights. \textcolor{black}{In particular, the closed-form solution for optimal power scaling is derived, revealing that the optimal magnitude is always obtained at its maximum value, which will be further corroborated through simulations.} To facilitate this analysis, we introduce two scaling coefficients $\delta_{r}> 0 $ and $\delta_{p} > 0$, such that $\mathbf{R}_{\mathbf{x}} = \delta_{r}\hat{\mathbf{R}}_{\mathbf{x}}$ and $\mathbf{p} = \sqrt{\delta_{p}}\hat{\mathbf{p}}$. Here, $\hat{\mathbf{R}}_{\mathbf{x}}$ and $\hat{\mathbf{p}}$ represent the feasible solutions to problem (P1) in the ISAC scenario and problem (P2) in the sensing-only scenario.   

\subsection{Sensing-Only Scenario}
Note that the objective function and the transmit power constraint \eqref{P1_I_cons2} at the IRS in problem (P2) depend on both transmit beamforming $\mathbf{R}_{\mathbf{x}}$ at the BS and amplification coefficients $\mathbf{p}$ at the active IRS. By letting $\hat{\mathbf{P}} = \mathrm{diag}(\hat{\mathbf{p}})$ and substituting $\mathbf{R}_{\mathbf{x}} = \delta_{r}\hat{\mathbf{R}}_{\mathbf{x}}$ and $\mathbf{P} = \sqrt{\delta_{p}}\hat{\mathbf{P}}$ into problem (P2), we obtain the following scaling coefficients optimization problem
 \begin{subequations}
	\begin{eqnarray}
	\!\!\!\!\!\!\!	(\text{P7}):	& \!\!\!\!{\mathop {\max}\limits_{\delta_{r} >0 \atop \delta_{p} >0 } }\!\!\!\!& \delta_{r} \delta_{p}^{2}  \nonumber\\
		& \!\!\!\!\text{s.t.}\!\!\!\!\!\! &\tilde{\beta}_{1}\delta_{r}\delta_{p}^{2}  + \tilde{\beta}_{2}\delta_{p}^{2}  +  \tilde{\beta}_{3}\delta_{r}\delta_{p}   + \tilde{\beta}_{4}\delta_{p} \leq P_{\text{s}}, \label{P8_cons1}\\
		&& \delta_{r}\leq \tilde{\beta}_{5},\!\!\! \label{P8_cons2}\\ 
		&& \delta_{p} \leq \tilde{\beta}_{6},\label{P8_cons3}
	\end{eqnarray}
\end{subequations}where $\tilde{\beta}_{1} = \mathrm{tr}\left(\hat{\mathbf{P}}^{2}\mathbf{E}\hat{\mathbf{P}}\mathbf{\Phi}\mathbf{G}\hat{\mathbf{R}}_{\mathbf{x}}\mathbf{G}^{H}\mathbf{\Phi}^{H}\hat{\mathbf{P}}\mathbf{E}^{H}\right)$, $\tilde{\beta}_{2} = \sigma_{\text{r}}^2\mathrm{tr}\left(\hat{\mathbf{P}}^{2}\mathbf{E}\hat{\mathbf{P}}^{2}\mathbf{E}^{H}\right)$, $\tilde{\beta}_{3} = \mathrm{tr}\left( \hat{\mathbf{P}}^{2}\mathbf{G}\hat{\mathbf{R}}_{\mathbf{x}}\mathbf{G}^{H} \right)$, $ \tilde{\beta}_{4} = 2\sigma_{\text{r}}^2 \mathrm{tr}\left(\hat{\mathbf{P}}^{2}\right)$, $\tilde{\beta}_{5} = \frac{ P_{\text{t}}}{\mathrm{tr}\left(\hat{\mathbf{R}}_{\mathbf{x}}\right)} $, and $\tilde{\beta}_{6} = \frac{a_{\text{max}}^{2}}{\max \left\{\hat{\mathbf{p}}^{2}\right\}}$. Note that the objective function in problem (P7) is a monotonically increasing function with respect to both $\delta_{r}$ and $\delta_{p}$. Let $\delta_{r}^{\star}$ and $\delta_{p}^{\star}$ denote the optimal solutions to problem (P7), respectively. Then, we have the following Proposition \ref{prop6} regarding the structure of the optimal solutions to problem (P7).
\begin{prop}\label{prop6}
	For sensing-only scenario, the relationship between $\delta_{r}$ and $\delta_{p}$ is derived as follows:
	\begin{itemize}
		\item With a given $\delta_{r}\leq \tilde{\beta}_{5}$, the optimal $\delta_{p}^{\star}$ to problem (P7) is given by $\min\left\{\tilde{\beta}_{6}, \Lambda_{1}\right\}$, where $\Lambda_{1} = \frac{\sqrt{\tilde{\beta}_{2}^{2}\delta_{r}^{2}+\left(2\tilde{\beta}_{2}\tilde{\beta}_{4}+4\tilde{\beta}_{1}P_{\text{s}}\right)\delta_{r} + \tilde{\beta}_{4}^{2} + 4\tilde{\beta}_{2}P_{\text{s}} }- \tilde{\beta}_{3}\delta_{r} - \tilde{\beta}_{4} }{2\tilde{\beta}_{1}\delta_{r} + 2\tilde{\beta}_{2} }$.
		\item With a given $\delta_{p}\leq \tilde{\beta}_{6}$, the optimal $\delta_{r}^{\star}$ to problem (P7) is given by $\min\left\{\tilde{\beta}_{5},\frac{P_{\text{s}}  - \tilde{\beta}_{2}\delta_{p}^{2} - \tilde{\beta}_{4}\delta_{p}}{\tilde{\beta}_{1}\delta_{p}^{2} + \tilde{\beta}_{3}\delta_{p}}\right\}$.
		\item If $\tilde{\beta}_{1}\tilde{\beta}_{5}\tilde{\beta}_{6}^{2}  + \tilde{\beta}_{2}\tilde{\beta}_{6}^{2}  +  \tilde{\beta}_{3}\tilde{\beta}_{5}\tilde{\beta}_{6}   + \tilde{\beta}_{4}\tilde{\beta}_{6} \leq P_{\text{s}}$, the optimal solutions to problem (P7) are given by $\delta_{r}^{\star} = \tilde{\beta}_{5}$ and $\delta_{p}^{\star} = \tilde{\beta}_{6}$.
	\end{itemize}
\end{prop}
\begin{IEEEproof}
Since the objective function in problem (P7) is a monotonically increasing function with respect to both $\delta_{r}$ and $\delta_{p}$, the optimal solutions to problem (P7) lie at the boundary of the feasible region. Therefore, given $\delta_{r} \leq \tilde{\beta}_{5}$, the optimal $\delta_{p}^{\star}$ is the minimum value between $\tilde{\beta}_{6}$ and the solution to the quadratic equation $\tilde{\beta}_{1}\delta_{r}\delta_{p}^{2} + \tilde{\beta}_{2}\delta_{p}^{2} + \tilde{\beta}_{3}\delta_{r}\delta_{p} + \tilde{\beta}_{4}\delta_{p} = P_{\text{s}}$. Similarly, the optimal $\delta_{r}^{\star}$ is obtained as described in Proposition \ref{prop6} for a given $\delta_{p} \leq \tilde{\beta}_{6}$.
\end{IEEEproof}

In fact, $\tilde{\beta}_{1}$ and $\tilde{\beta}_{2}$ are significantly smaller than $\tilde{\beta}_{3}$ and $\tilde{\beta}_{4}$ due to the fact that the target response matrix $\mathbf{E}$ and the noise variance $\sigma_{\text{r}}^{2}$ typically exhibit small values. Consequently, problem (P7) can be approximated as problem (P7.1) by neglecting the terms that contain $\tilde{\beta}_{1}$ and $\tilde{\beta}_{2}$.
 \begin{subequations}
	\begin{eqnarray}
			\!\!\!\!\!\!\!	(\text{P7.1}):	& \!\!\!\!{\mathop {\max}\limits_{\delta_{r} >0 \atop \delta_{p} >0 } }\!\!\!\!& \delta_{r} \delta_{p}^{2}  \nonumber\\
		& \!\!\!\!\text{s.t.}\!\!\!\!\!\! & \tilde{\beta}_{3}\delta_{r}\delta_{p}   + \tilde{\beta}_{4}\delta_{p} \leq P_{\text{s}}, \label{P8_1_cons1}\\
		&& \eqref{P8_cons2}-\eqref{P8_cons3}.\nonumber
	\end{eqnarray}
\end{subequations}
When $\tilde{\beta}_{3}\tilde{\beta}_{5}\tilde{\beta}_{6}   + \tilde{\beta}_{4}\tilde{\beta}_{6} > P_{\text{s}}$, the optimal solution to problem (P7) should satisfy $\tilde{\beta}_{3}\delta_{r}\delta_{p}   + \tilde{\beta}_{4}\delta_{p} = P_{\text{s}}$. Hence, we have $ \delta_{r} = \frac{ P_{\text{s}} - \tilde{\beta}_{4}\delta_{p}}{ \tilde{\beta}_{3} \delta_{p}}$. Then, by substituting it into problem (P7.1), the objective function is rewritten as $\frac{1}{ \tilde{\beta}_{3}} \left(P_{\text{s}} - \tilde{\beta}_{4}\delta_{p} \right) \delta_{p}$. Therefore, the objective function in problem (P7.1) is monotonically increasing with $\delta_{p} \leq \frac{P_{\text{s}}}{2\tilde{\beta}_{4}}$ and decreasing with $\delta_{p} > \frac{P_{\text{s}}}{2\tilde{\beta}_{4}}$. Let $\delta_{p}^{\diamond}$ denote the optimal $\delta_{p}$ to problem (P7.1). As a result, we have the following results:
\begin{prop}
	If   $\tilde{\beta}_{3}\tilde{\beta}_{5}\tilde{\beta}_{6}   + \tilde{\beta}_{4}\tilde{\beta}_{6} > P_{\text{s}}$, the optimal solution $\delta_{p}^{\diamond}$ to problem (P7.1) is given by
	 \begin{align}
	 	\delta_{p}^{\diamond} = \left\{\begin{array}{ll}
	 		\tilde{\beta}_{6}, &\frac{P_{\text{s}}}{2\tilde{\beta}_{4}} \geq \tilde{\beta}_{6}, \\
	 		\frac{P_{\text{s}}}{2\tilde{\beta}_{4}}, &\frac{P_{\text{s}}}{2\tilde{\beta}_{4}} < \tilde{\beta}_{6}.
	 	\end{array} \right. \label{optimal_delat_p}
	 \end{align}
\end{prop}
\begin{IEEEproof}
	The proof is similar to that of Proposition \ref{prop6}.
\end{IEEEproof}
\begin{rem}
The condition $\frac{P_{\text{s}}}{2\tilde{\beta}_4} \geq \tilde{\beta}_6$ is equivalent to $\frac{ P_{\text{s}} \max \left\{\hat{\mathbf{p}}^2\right\}}{4\sigma_{\text{r}}^2 \mathrm{tr}\left(\hat{\mathbf{P}}^2\right) a_{\text{max}}^2} \geq 1$. This indicates that a large $P_{\text{s}}$ and a small $a_{\text{max}}$ are more likely to result in the optimal solution $\delta_{p}^{\diamond}$ to problem (P7.1) being located at $\tilde{\beta}_6$, implying the maximum value of the elements in the optimal $\mathbf{p}$ equals $a_{\text{max}}$. In practical communication systems, there is a high probability that $\frac{P_{\text{s}} \max \left\{\hat{\mathbf{p}}^2\right\}}{4\sigma_{\text{r}}^2 \mathrm{tr}\left(\hat{\mathbf{P}}^2\right) a_{\text{max}}^2} \geq 1$ holds as the noise variance $\sigma_{\text{r}}^2$ is typically rather small \cite{bjornson2024introduction}. 
\end{rem}

\subsection{ISAC Scenario}
According to $\mathbf{R}_{\mathbf{x}} = \delta_{r}\hat{\mathbf{R}}_{\mathbf{x}}$, we have $\mathbf{R}_{0} = \delta_{r}\hat{\mathbf{R}}_{0}$ and $\mathbf{w}_{k} = \sqrt{\delta_{r}}\hat{\mathbf{w}}_{k}, \forall k \in \mathcal{K}$, where $\hat{\mathbf{R}}_{0}$ and $\{\hat{\mathbf{w}}_{k}\}$ are given feasible solutions to problem (P1). Similar to problem (P7), the  optimization of the scaling coefficients for the ISAC scenario is formulated as
\begin{subequations}
	\begin{eqnarray}
			\!\!\!\!\!\!\!	(\text{P8}):	& \!\!\!\!{\mathop {\max}\limits_{\delta_{r} >0 \atop \delta_{p} >0 } }\!\!\!\!& \delta_{r} \delta_{p}^{2}  \nonumber\\
		& \!\!\!\!\text{s.t.}\!\!\!\!\!\! &\eqref{P8_cons1}-\eqref{P8_cons3},\nonumber\\
		&& \tilde{\beta}_{7}\delta_{r}\delta_{p} + \tilde{\beta}_{8}\delta_{p} + \sigma_{\text{u}}^2 \leq 0,\label{P9_cons4}
	\end{eqnarray}
\end{subequations}
in which
\begin{align}
	\tilde{\beta}_{7} &= \sum\nolimits_{k' \in \mathcal{K} \backslash \{k\}}\!\!\!\!\mathrm{tr}\! \left( \mathbf{\Phi}\breve{\mathbf{H}}_{k}^{H}\mathbf{G}\hat{\mathbf{W}}_{k'}\mathbf{G}^{H}\breve{\mathbf{H}}_{k}\mathbf{\Phi}^{H}\hat{\bar{\mathbf{P}}}\right) \nonumber\\
	&\quad+\mathrm{tr}\! \left(\mathbf{\Phi}\breve{\mathbf{H}}_{k}^{H}\mathbf{G} \hat{\mathbf{R}}_{0}\mathbf{G}^{H}\breve{\mathbf{H}}_{k}\mathbf{\Phi}^{H}\bar{\mathbf{P}}\right) \nonumber\\
	&\quad-\frac{1}{\Gamma_{k}}\mathrm{tr}\! \left(\mathbf{\Phi} \breve{\mathbf{H}}_{k}^{H}\mathbf{G}\hat{\mathbf{W}}_{k}\mathbf{G}^{H}\breve{\mathbf{H}}_{k}\mathbf{\Phi}^{H}\hat{\bar{\mathbf{P}}}\right),\\
	\tilde{\beta}_{8} &= \sigma_{\text{r}}^2\mathrm{tr} \left(\mathbf{H}_{k}\hat{\bar{\mathbf{P}}}\right), 
\end{align}
with $\hat{\bar{\mathbf{P}}} = \hat{\mathbf{p}}\hat{\mathbf{p}}^{T}$. Because the objective function in (P8) is a monotonically increasing function with respect to $\delta_{r}$ and $\delta_{p}$, we have the following proposition.
\begin{prop}\label{prop8}
		For the ISAC scenario, the relationship between $\delta_{r}$ and $\delta_{p}$ is derived as follows:
	\begin{itemize}
		\item With a given $\delta_{r}\leq \tilde{\beta}_{5}$, the optimal $\delta_{p}^{\star}$ to problem (P8) is given by $\min\left\{\tilde{\beta}_{6},\Lambda_{1}\right\}$.
		\item With a given $\delta_{p}\leq \tilde{\beta}_{6}$, the optimal $\delta_{r}^{\star}$ to problem (P8) is given by $\min \left\{\tilde{\beta}_{5},\frac{P_{\text{s}}  - \tilde{\beta}_{2}\delta_{p}^{2} - \tilde{\beta}_{4}\delta_{p}}{\tilde{\beta}_{1}\delta_{p}^{2} + \tilde{\beta}_{3}\delta_{p}}\right\}$.
		\item If $\tilde{\beta}_{1}\tilde{\beta}_{5}\tilde{\beta}_{6}^{2}  + \tilde{\beta}_{2}\tilde{\beta}_{6}^{2}  +  \tilde{\beta}_{3}\tilde{\beta}_{5}\tilde{\beta}_{6}   + \tilde{\beta}_{4}\tilde{\beta}_{6} \leq P_{\text{s}}$ and $\tilde{\beta}_{5}\tilde{\beta}_{6}\tilde{\beta}_{7} + \tilde{\beta}_{6}\tilde{\beta}_{8} + \sigma_{\text{u}}^2 \leq 0$ hold, the optimal solution to problem (P8) are $\delta_{r}^{\star} = \tilde{\beta}_{5}$ and $\delta_{p}^{\star} = \tilde{\beta}_{6}$.
	\end{itemize}
\end{prop}
\begin{IEEEproof}
The proof is similar to that of Proposition \ref{prop6}.	
\end{IEEEproof}

By comparing Propositions \ref{prop6} and \ref{prop8}, we observe that the structures of the optimal scaling coefficients, $\delta_{r}$ and $\delta_{p}$, for the ISAC scenario are identical as those in the sensing-only scenario. Consequently, the optimal scaling coefficients always locate at the boundary of region defined by the constraints in \eqref{P8_cons1}-\eqref{P8_cons3}.
\vspace{-2mm}
\section{Numerical Results}
This section provides numerical results to validate the effectiveness of our proposed AO-based design. In the simulation, we adopt the Rician fading channel model with the K-factor being $5$ dB for channels between the BS and the IRS. The channels between the IRS and the target are assumed to be LoS channels. Additionally, we set the noise power as $\sigma_{\text{r}}^{2}= \sigma_{\text{b}}^{2} =\sigma_{\text{u}}^2= -110$ dBm.
In particular, the BS and active IRS are located at $(0,0)$ meters~(m) and $(0,25)$ m, respectively. The CUs are located at $(-20,10)$ m and $(20,10)$ m. The target is located at $(-15,10)$ m.  We also set the SINR constraints to be $\Gamma_{k} = \Gamma, \forall  k \in \mathcal{K}$, and the ISAC dwell time as $T = 100$. To better illustrate the superiority of the proposed AO-based design, we adopt the following benchmarks for comparison.
\begin{itemize}
\item \textbf{Transmit beamforming (BF) only}: The active IRS implements random reflection coefficients as $\bm{\psi} = a_{\text{max}}e^{\jmath\mathrm{arg}\left\{\mathbf{r}\right\}}$, where ${\mathbf{r}} \sim \mathcal{CN}\left(\mathbf{0},\mathbf{I}_N\right)$. Accordingly, we only optimize the transmit beamforming at the BS by solving problems (P3) and (P5) in Sections III-A and IV-A, respectively.

\item \textbf{Reflective BF only}: This benchmark is for the sensing-only scenario, in which the BS adopts the isotropic transmission by setting $\mathbf{R}_{\mathbf{x}}=\frac{P_{\text{t}}}{M}\mathbf{I}_{M}$. Then, we optimize the reflective beamforming at the IRS by solving problems (P4) in Sections III-B.

\item \textbf{Passive IRS-enabled design}: This benchmark refers to passive IRS deployment. In this scheme, the CRB only depends on the transmit beamforming. For the sensing-only scenario, we only optimize the transmit beamforming $\mathbf{R}_{\mathbf{x}}$ at the BS \cite{song2023intelligent}. For the  ISAC scenario, we optimize transmit beamforming  $\mathbf{R}_{\mathbf{x}}$ at the BS and $\mathbf{\Phi}$ at the IRS according to the scheme in Section IV-B-1.

\item \textbf{ZF-BF}: This benchmark is designed for the ISAC scenario. First, we design the transmit information beamformers based on the ZF principle. \textcolor{black}{In particular, by defining $\tilde{\mathbf{H}} = [\bar{\mathbf{h}}_{1},\ldots,\bar{\mathbf{h}}_{K}]$} and   $\tilde{\mathbf{W}}^{\text{ZF}} = \tilde{\mathbf{H}}(\tilde{\mathbf{H}}^{H} \tilde{\mathbf{H}})^{-1}$, we set the ZF beamformer for each CU $k$ as $\mathbf{w}_{k}^{\text{ZF}} =  \sqrt{p_{k}}{\tilde{\mathbf{w}}_{k}^{\text{ZF}}}/{\|\tilde{\mathbf{w}}_{k}^{\text{ZF}}\|}$, where $\tilde{\mathbf{w}}_{k}^{\text{ZF}}$ denotes the $k$-th column of $\tilde{\mathbf{W}}^{\text{ZF}}$ and $p_{k}$ is the transmit power for CU $k$. Then, we optimize the transmit  power $\{p_{k}\}$, the sensing beamformers $\mathbf{R}_0$, and the reflective beamformers $\mathbf{\Psi}$ by AO, similarly as presented in Section IV. 
\end{itemize}
\subsection{Sensing-Only Scenario}
\begin{figure}[tbp]
	\setlength{\abovecaptionskip}{2pt}
	\setlength{\belowcaptionskip}{-14pt}
	\centering
	\includegraphics[width= 0.38\textwidth]{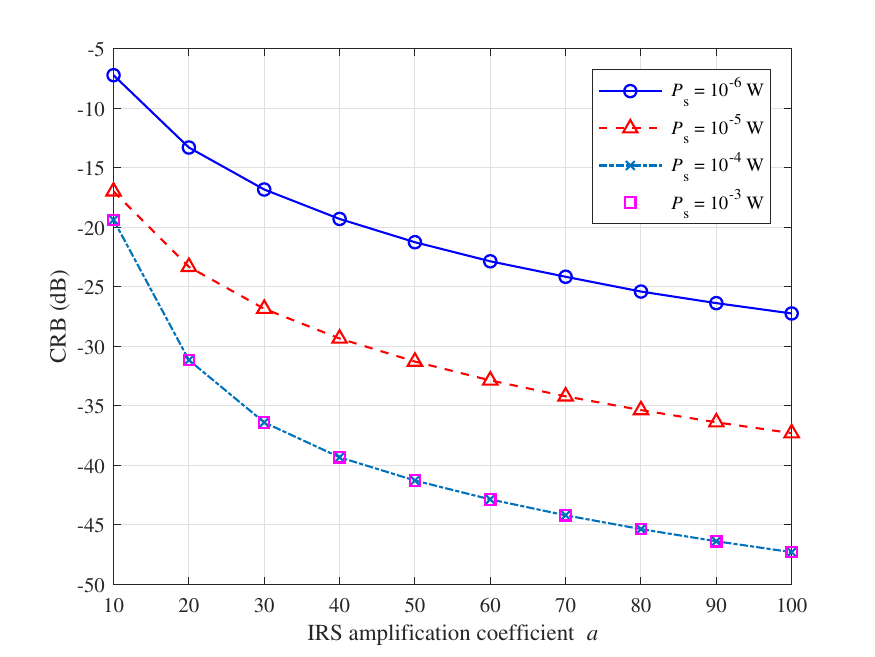}
	\DeclareGraphicsExtensions.
	\caption{\color{black}The achieved CRB versus the amplification coefficients $a_{n} = a, \forall n\in \mathcal{N}$ at IRS for the sensing-only scenario with $P_{\text{t}} = 50$ W and $M=N=8$.}
	\label{CRBVsIRSa}
\end{figure}
First, we consider sensing-only scenario. Fig.~\ref{CRBVsIRSa} shows the achieved CRB versus the amplification coefficients $a_{n} = a, \forall n \in \mathcal{N}$ at the IRS for the sensing-only scenario, where we set $P_{\text{t}} = 50$ W and the reflective beamforming is $\bm{\psi} = a e^{\jmath\mathrm{arg}\left\{\mathbf{r}\right\}}$ with ${\mathbf{r}} \sim \mathcal{CN}\left(\mathbf{0},\mathbf{I}_N\right)$. It is observed that the achieved CRB decreases as the amplification coefficients $a$ increases. This indicates that the optimal amplification coefficient always tends to approach the maximum amplification gain at the IRS, which coincides with the scaling law analysis in Section V. Additionally, it is also observed that the higher the maximum transmit power $P_{\text{s}}$ at the IRS, the better sensing performance can be achieved. This is because the maximum transmit power constraint \eqref{IRS_power_cons1} at the IRS limits the transmit beamforming at the BS.

\begin{figure}[tbp]
	\setlength{\abovecaptionskip}{2pt}
	\setlength{\belowcaptionskip}{-12pt}
	\centering
	\includegraphics[width= 0.38\textwidth]{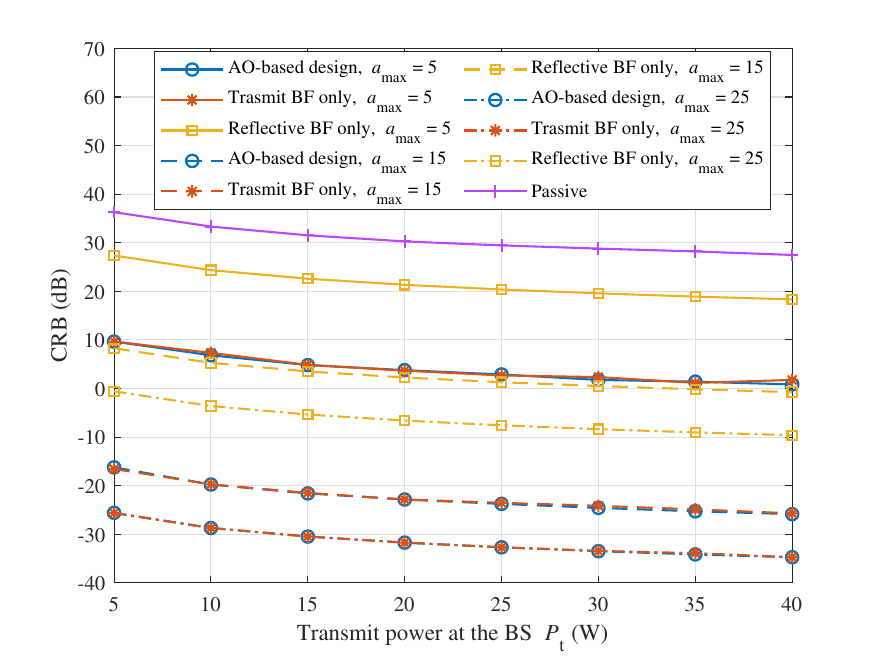}
	\DeclareGraphicsExtensions.
	\caption{\color{black}The achieved CRB versus the maximum transmit power $P_{\text{t}}$ at the BS for sensing-only scenario with $P_{\text{s}} = 0.01$ W and $M=N=8$.}
	\label{CRBVsPt}
\end{figure}
Fig.~\ref{CRBVsPt} plots the achieved CRB versus the maximum transmit power $P_{\text{t}}$ at the BS. It is observed that our proposed AO-based design outperforms the `reflective BF only' and passive IRS schemes over the whole regime of $P_{\text{t}}$, which shows the benefit of active IRS deployment. Besides, the `transmit BF only' design performs close to the proposed AO-based design. This indicates the optimal magnitude of the reflection coefficients are always equal to the maximum amplification gain, while the phase of these coefficients has a negligible effect on the CRB. \textcolor{black}{\it Thus, merely optimizing the transmit beamforming at the BS with the maximum amplification gain at the active IRS always achieves near-optimal sensing performance.} Furthermore, it is observed that a higher maximum amplification gain at the IRS leads to improved sensing performance. This is because a more relaxed constraint on the maximum amplification gain of the IRS elements provides greater degrees of freedom for reflective beamforming.

\begin{figure}[tbp]
	\setlength{\abovecaptionskip}{2pt}
	\setlength{\belowcaptionskip}{-14pt}
	\centering
	\includegraphics[width= 0.38\textwidth]{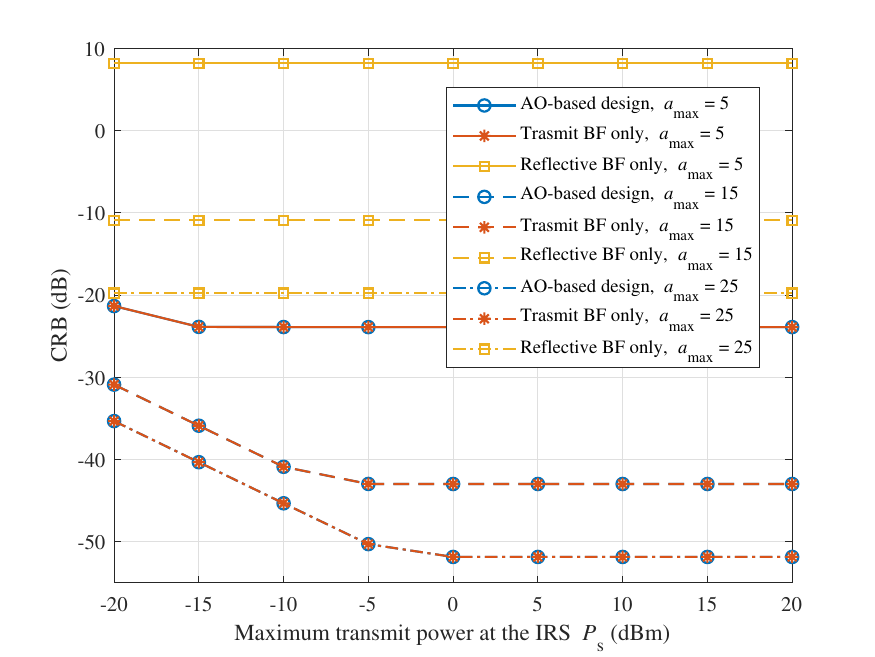}
	\DeclareGraphicsExtensions.
	\caption{\color{black}The achieved CRB versus the maximum transmit power $P_{\text{s}}$ at the IRS for sensing-only scenario with $P_{\text{t}} = 40$ W and $M=N=8$.}
	\label{CRBVsPs}
\end{figure}
Fig.~\ref{CRBVsPs} shows the achieved CRB versus the maximum transmit power $P_{\text{s}}$ at the IRS for the sensing-only scenario. In the low $P_{\text{s}}$ regime, it is observed that the CRB decreases as $P_{\text{s}}$ increases. This phenomenon occurs because the received echo signal power is primarily constrained by the maximum transmit power budget at the IRS. Subsequently, in the high $P_{\text{s}}$ regime, it is observed that the CRBs remain constant. This is attributed to the fact that the maximum transmit power budget at the IRS is large enough, and the received echo signal power is mainly constrained by the maximum amplification gain instead. Moreover, the `reflective BF only' scheme remains constant in the region of $P_{\text{s}}$. This is due to the maximum transmit power budget at the IRS, i.e., the constraint \eqref{P1_I_cons2} is always satisfied and the optimal solution to problem (P4) satisfies Proposition \ref{prop4}.
\subsection{ISAC Scenario}
\begin{figure}[tbp]
	\setlength{\abovecaptionskip}{2pt}
	\setlength{\belowcaptionskip}{-12pt}
	\centering
	\includegraphics[width= 0.38\textwidth]{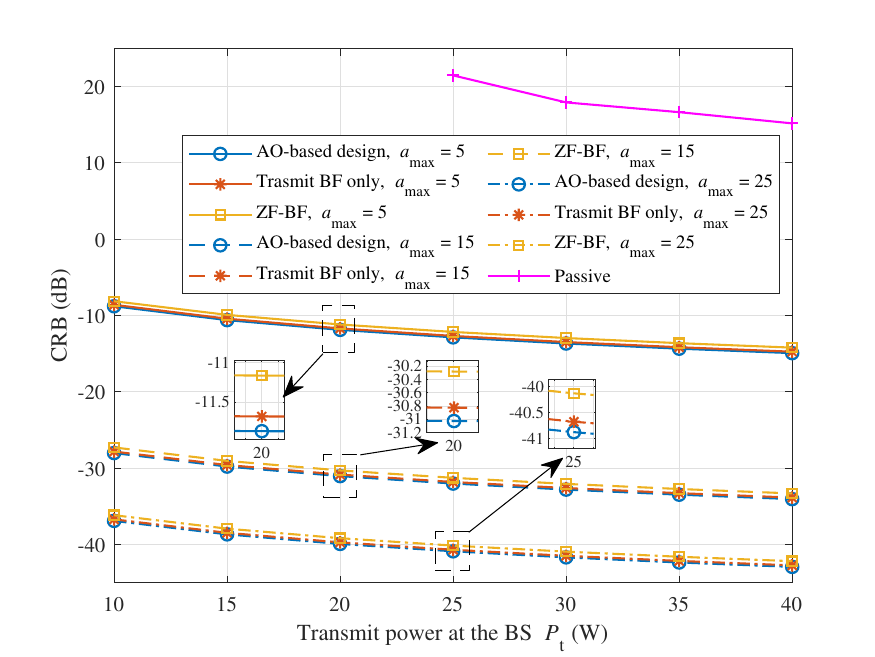}
	\DeclareGraphicsExtensions.
	\caption{\color{black}The achieved CRB versus the maximum transmit power $P_{\text{t}}$ at the BS for the ISAC scenario with $P_{\text{s}} = 0.01$ W and $M=N=8$.}
	\label{CRBVsPtISAC}
\end{figure}
Next, we consider the ISAC scenario. Fig.~\ref{CRBVsPtISAC} shows the achieved CRB versus the maximum transmit power $P_{\text{t}}$ at the BS, where we set $\Gamma = 10$ dB. It is observed that our proposed AO-based design outperforms the `transmit BF only' and `ZF-BF' schemes, and the CRB performance achieved by active IRS outperforms that of passive IRS by a significant margin. As the maximum transmit power $P_{\text{t}}$ increases, the CRB performances by our proposed AO-based design and the other benchmark schemes gradually approach a constant value. This is due to the fact that the CRB performance mainly depends on the maximum transmit power budget and the maximum amplification gain at the IRS when the maximum transmit power at the BS is high. It is also observed that the higher the maximum amplification gain at the IRS, the system can achieve a better CRB performance. This indicates that the performance gain of the active IRS compared to the passive IRS highly relies on the maximum amplification gain. 

\begin{figure}[tbp]
	\setlength{\abovecaptionskip}{2pt}
	\setlength{\belowcaptionskip}{-14pt}
	\centering
	\includegraphics[width= 0.40\textwidth]{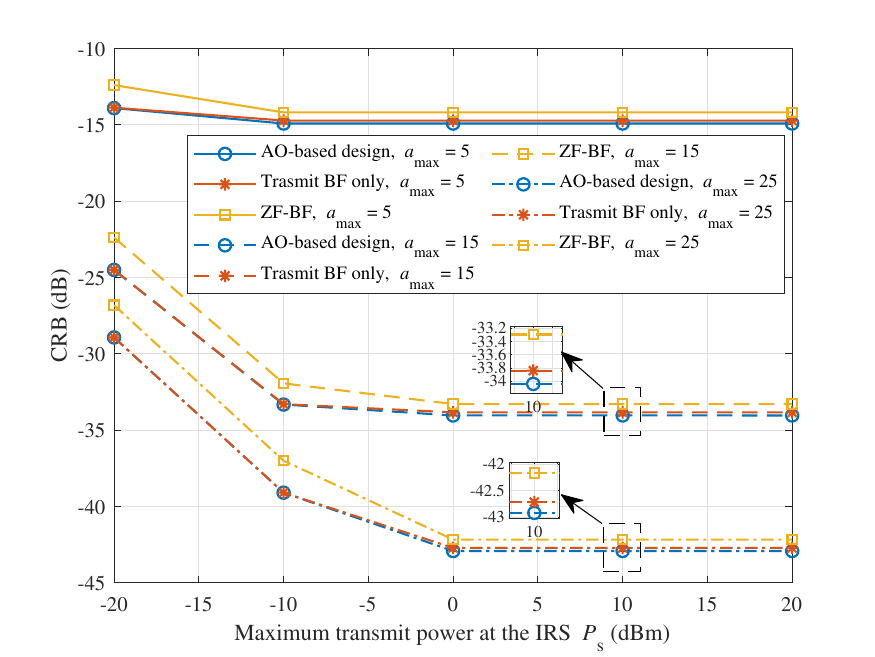}
	\DeclareGraphicsExtensions.
	\caption{\color{black}The achieved CRB versus the maximum transmit power $P_{\text{s}}$ at the IRS for the ISAC scenario with $P_{\text{t}} = 40$ W and $M=N=8$.}
	\label{CRBVsPsISAC}
\end{figure}
Fig.~\ref{CRBVsPsISAC} shows the achieved CRB versus the maximum transmit power $P_{\text{s}}$ at the IRS, where we set $\Gamma = 10$ dB. It is observed that in the low $P_{\text{s}}$ regime, the CRB achieved by our proposed AO-based design is close to that by the `transmit BF only' scheme and decreases as $P_{\text{s}}$ increases. This is because the transmit beamforming at the BS is constrained by the maximum transmit power limit at the IRS, as described in \eqref{IRS_power_cons1}, and under such conditions, the sensing performance primarily depends on the beamforming at the BS. However, as $P_{\text{s}}$ increases, the transmit power constraint at the IRS no longer restricts the transmit beamforming quality at the BS, allowing the benefits of reflective beamforming to gradually become more pronounced. It is also observed that the `ZF-BF' scheme performs worst among all schemes. This indicates the importance of transmit beamforming optimization at the BS for minimizing the achieved CRB. 

\begin{figure}[tbp]
	\setlength{\abovecaptionskip}{2pt}
	\setlength{\belowcaptionskip}{-15pt}
	\centering
	\includegraphics[width= 0.40\textwidth]{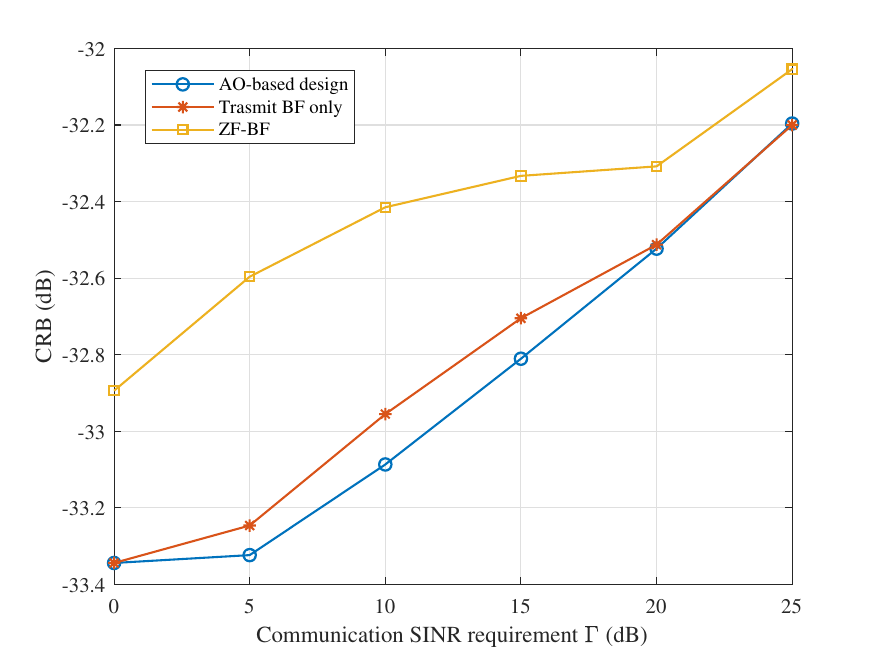}
	\DeclareGraphicsExtensions.
	\caption{\color{black}The achieved CRB versus the SINR requirements $\Gamma$ at the CUs for the ISAC scenario with $P_{\text{t}} = 20$ W, $P_{\text{s}} = 0.01$ W, $a_{\text{max}}=15$, and $M=N=8$.}
	\label{CRBVsSinrISAC}
\end{figure}
Fig.~\ref{CRBVsSinrISAC} shows the achieved CRB versus the communication SINR constraint $\Gamma$ at each CU. It is observed that our proposed AO-based design achieves a larger achievable CRB-SINR region compared to benchmark schemes. With extremely low and high SINR requirements, the CRB performance of the proposed AO-based design and `transmit BF only' design is close. This is because that most of transmit power at the BS is allocated for sensing at the low SINR requirement and for communication at the high SINR requirement. In these cases, we only need to optimize the transmit beamforming at the BS. 
It is also observed that the CRB performance of the `ZF-BF' scheme is the worst among all schemes. This observation suggests that ZF beamforming at the BS is suboptimal for the ISAC scenario. The ZF beamforming primarily focuses on mitigating inter-user interference, neglecting the enhancement of the signal strength directed towards the target, which is crucial for effective sensing.
\vspace{-2mm}
\section{Conclusion}
\vspace{-1mm}
This paper investigated the active IRS-enabled ISAC with extended target, where an active IRS is deployed to assist the BS providing ISAC service. \textcolor{black}{In particular, we derived the closed-form CRB for estimating the target response matrix of an extended target. It demonstrated that the CRB depends solely on the transmit beamforming at the BS and the amplification gain at the active IRS. In order to minimize the estimation CRB to enhance sensing performance, we derived a closed-form optimal transmit beamforming under specific conditions for the sensing-only scenario, which helped reduce computational complexity. Subsequently, we proposed efficient algorithms to jointly optimize the transmit beamforming at the BS, the phase shifts and the amplification at the IRS based on SDR and SCA techniques for the ISAC scenario. In addition, we analyzed the scaling laws of transmit beamforming at the BS and amplification coefficients at the active IRS, establishing a relationship between their optimal scaling and confirming that the active IRS always uses maximum amplification gain for practical wireless system settings.} Finally, numerical results demonstrated the effectiveness of our proposed AO-based design, showcasing the advantages of active IRS over passive IRS configurations and revealing superior performance compared to various benchmark schemes.

\allowdisplaybreaks 
\begin{appendices}
	\section{Proof of Proposition \ref{prop_FIM_bs}}\label{prop_FIM1_proof}
	We first decompose the estimation parameter vector $\bm{\xi}$  into real and imaginary parts that are denoted by $\Re \{ {\bm{\xi}}\}$ and $\Im \{ {\bm{\xi}}\}$, respectively. Then, each part in the FIM \eqref{FIM_def_bs} are derived as
	{\small\begin{align}
		&\!\frac{{\partial \bm{\eta }}}{{\partial \Re \{ {\bm{\xi}}\} }} \!=\! \frac{{\partial \left( {{{\bf{X}}^T}{\mathbf{G}^T}{\mathbf{\Psi }}\! \otimes \!{\mathbf{G}^T}{\mathbf{\Psi }}} \right){\bm{\xi}}}}{{\partial \Re \{ {\bm{\xi}}\} }} \!= \left( {{{\bf{X}}^T}{\mathbf{G}^T}{\mathbf{\Psi }} \otimes {\mathbf{G}^T}{\mathbf{\Psi }}} \right),\\
		&\!\frac{{\partial \bm{\eta }}}{{\partial \Im \{ {\bm{\xi}}\} }} \!=\! \frac{{\partial \left( {{{\bf{X}}^T}{\mathbf{G}^T}{\mathbf{\Psi }} \!\otimes\! {\mathbf{G}^T}{\mathbf{\Psi }}} \right){\bm{\xi}}}}{{\partial \Im \{ {\bm{\xi}}\} }} \!= j\left( {{{\bf{X}}^T}{\mathbf{G}^T}\!{\mathbf{\Psi }} \! \otimes \! {\mathbf{G}^T}{\mathbf{\Psi }}} \right).
	\end{align}}Thus, we have 
	{\small\begin{align}
	&\Re \left( {\frac{{\partial \mathbf{\eta }_{}^H}}{{\partial \Re \{ {\bm{\xi}}\} }}{\bf{R}}_{\bf{y}}^{ - 1}\frac{{\partial {\mathbf{\eta }_{}}}}{{\partial \Re \{ {\bm{\xi}}\} }}} \right) \nonumber\\
	&\!=\! \Re \left\{ {\left( {{{\mathbf{\Psi }}^H}{\mathbf{G}^*}{{\bf{X}}^*} \!\otimes \!{{\mathbf{\Psi }}^H}{\mathbf{G}^*}} \right){\bf{R}}_{\bf{y}}^{ - 1}\left( {{{\bf{X}}^T}{\mathbf{G}^T}{\mathbf{\Psi }} \otimes {\mathbf{G}^T}{\mathbf{\Psi }}} \right)} \right\},\\
	&\Re \left( {\frac{{\partial \mathbf{\eta }_{}^H}}{{\partial \Re \{ {\bm{\xi}}\} }}{\bf{R}}_{\bf{y}}^{ - 1}\frac{{\partial {\mathbf{\eta }_{}}}}{{\partial \Im \{ {\bm{\xi}}\} }}} \right) \nonumber\\
	&\!=\!  - \Im \left\{ \!{\left( {{{\mathbf{\Psi }}^H}{\mathbf{G}^*}{{\bf{X}}^*} \!\otimes \!{{\mathbf{\Psi }}^H}{\mathbf{G}^*}} \right){\bf{R}}_{\bf{y}}^{ - 1}\left( {{{\bf{X}}^T}{\mathbf{G}^T}{\mathbf{\Psi }} \!\otimes \!{\mathbf{G}^T}{\mathbf{\Psi }}} \right)} \!\right\},
	\end{align}}which completes the proof of Proposition \ref{prop_FIM_bs}.
\end{appendices}
\bibliographystyle{IEEEtran}
\bibliography{IEEEabrv,myref}

\begin{thebibliography}{10}
\providecommand{\url}[1]{#1}
\csname url@samestyle\endcsname
\providecommand{\newblock}{\relax}
\providecommand{\bibinfo}[2]{#2}
\providecommand{\BIBentrySTDinterwordspacing}{\spaceskip=0pt\relax}
\providecommand{\BIBentryALTinterwordstretchfactor}{4}
\providecommand{\BIBentryALTinterwordspacing}{\spaceskip=\fontdimen2\font plus
\BIBentryALTinterwordstretchfactor\fontdimen3\font minus
  \fontdimen4\font\relax}
\providecommand{\BIBforeignlanguage}[2]{{%
\expandafter\ifx\csname l@#1\endcsname\relax
\typeout{** WARNING: IEEEtran.bst: No hyphenation pattern has been}%
\typeout{** loaded for the language `#1'. Using the pattern for}%
\typeout{** the default language instead.}%
\else
\language=\csname l@#1\endcsname
\fi
#2}}
\providecommand{\BIBdecl}{\relax}
\BIBdecl

\bibitem{cui2021integrating}
Y.~Cui, F.~Liu, X.~Jing, and J.~Mu, ``{Integrating sensing and communications
  for ubiquitous IoT: Applications, trends, and challenges},'' \emph{IEEE
  Netw.}, vol.~35, no.~5, pp. 158--167, Sep. 2021.

\bibitem{zhang2021overview}
J.~A. Zhang, F.~Liu, C.~Masouros, R.~W. Heath, Z.~Feng, L.~Zheng, and
  A.~Petropulu, ``{An overview of signal processing techniques for joint
  communication and radar sensing},'' \emph{IEEE J. Sel. Topics Signal
  Process.}, vol.~15, no.~6, pp. 1295--1315, Nov. 2021.

\bibitem{liu2022integrated}
F.~Liu, Y.~Cui, C.~Masouros, J.~Xu, T.~X. Han, Y.~C. Eldar, and S.~Buzzi,
  ``{Integrated sensing and communications: Towards dual-functional wireless
  networks for 6G and beyond},'' \emph{IEEE J. Sel. Areas Commun.}, vol.~40,
  no.~6, pp. 1728--1767, Jun. 2022.

\bibitem{huang2019reconfigurable}
C.~Huang, A.~Zappone, G.~C. Alexandropoulos, M.~Debbah, and C.~Yuen,
  ``{Reconfigurable intelligent surfaces for energy efficiency in wireless
  communication},'' \emph{IEEE Trans. Wireless Commun.}, vol.~18, no.~8, pp.
  4157--4170, Aug. 2019.

\bibitem{hua2022joint}
M.~Hua, Q.~Wu, C.~He, S.~Ma, and W.~Chen, ``{Joint active and passive
  beamforming design for IRS-aided radar-communication},'' \emph{IEEE Trans.
  Wireless Commun.}, vol.~22, no.~4, pp. 2278--2294, Apr. 2023.

\bibitem{hua2023intelligent}
M.~Hua, Q.~Wu, W.~Chen, Z.~Fei, H.~C. So, and C.~Yuen, ``{Intelligent
  reflecting surface assisted localization: Performance analysis and algorithm
  design},'' \emph{IEEE Wireless Commun. Lett.}, vol.~13, no.~1, pp. 84--88,
  Jan. 2023.

\bibitem{di2020smart}
M.~Di~Renzo, A.~Zappone, M.~Debbah, M.-S. Alouini, C.~Yuen, J.~De~Rosny, and
  S.~Tretyakov, ``{Smart radio environments empowered by reconfigurable
  intelligent surfaces: How it works, state of research, and the road ahead},''
  \emph{IEEE J. Sel. Areas Commun.}, vol.~38, no.~11, pp. 2450--2525, Nov.
  2020.

\bibitem{wu2021intelligent}
Q.~Wu, S.~Zhang, B.~Zheng, C.~You, and R.~Zhang, ``{Intelligent reflecting
  surface-aided wireless communications: A tutorial},'' \emph{IEEE Trans.
  Commun.}, vol.~69, no.~5, pp. 3313--3351, May 2021.

\bibitem{yu2021irs}
X.~Yu, D.~Xu, D.~W.~K. Ng, and R.~Schober, ``{IRS-assisted green communication
  systems: Provable convergence and robust optimization},'' \emph{IEEE Trans.
  Commun.}, vol.~69, no.~9, pp. 6313--6329, Sep. 2021.

\bibitem{wu2023intelligent}
Q.~Wu, T.~Q. Duong, D.~W.~K. Ng, R.~Schober, and R.~Zhang, \emph{{Intelligent
  Surfaces Empowered 6G Wireless Network}}.\hskip 1em plus 0.5em minus
  0.4em\relax John Wiley \& Sons, 2023.

\bibitem{xu2020resource}
D.~Xu, X.~Yu, Y.~Sun, D.~W.~K. Ng, and R.~Schober, ``{Resource allocation for
  IRS-assisted full-duplex cognitive radio systems},'' \emph{IEEE Trans.
  Commun.}, vol.~68, no.~12, pp. 7376--7394, Dec. 2020.

\bibitem{song2024cramer}
X.~Song, X.~Qin, J.~Xu, and R.~Zhang, ``{Cram{\'e}r-Rao bound minimization for
  IRS-enabled multiuser integrated sensing and communications},'' \emph{IEEE
  Trans. Wireless Commun.}, vol.~23, no.~8, pp. 9714--9729, Aug. 2024.

\bibitem{Xianxin2025overview}
X.~Song, Y.~Fang, F.~Wang, Z.~Ren, X.~Yu, Y.~Zhang, F.~Liu, J.~Xu, D.~W.~K. Ng,
  R.~Zhang, and S.~Cui, ``{An overview on IRS-enabled sensing and
  communications for 6G: architectures, fundamental limits, and joint
  beamforming designs},'' \emph{Sci. China Inf. Sci.}, vol.~68, no.~5, p.
  150301, May 2025.

\bibitem{jiang2021intelligent}
Z.-M. Jiang, M.~Rihan, P.~Zhang, L.~Huang, Q.~Deng, J.~Zhang, and E.~M.
  Mohamed, ``Intelligent reflecting surface aided dual-function radar and
  communication system,'' \emph{IEEE Sys. J.}, vol.~16, no.~1, pp. 475--486,
  Mar. 2021.

\bibitem{yan2022reconfigurable}
S.~Yan, S.~Cai, W.~Xia, J.~Zhang, and S.~Xia, ``A reconfigurable intelligent
  surface aided dual-function radar and communication system,'' in \emph{Proc.
  IEEE Int. Symp. Joint Commun. \& Sensing (JC\&S)}, Seefeld, Austria, Mar.
  2022, pp. 1--6.

\bibitem{liu2022joint}
R.~Liu, M.~Li, Y.~Liu, Q.~Wu, and Q.~Liu, ``{Joint transmit waveform and
  passive beamforming design for RIS-aided DFRC systems},'' \emph{IEEE J. Sel.
  Topics Signal Process.}, vol.~16, no.~5, pp. 995--1010, Aug. 2022.

\bibitem{wang2021joint1}
X.~Wang, Z.~Fei, Z.~Zheng, and J.~Guo, ``{Joint waveform design and passive
  beamforming for RIS-assisted dual-functional radar-communication system},''
  \emph{IEEE Trans. Veh. Technol.}, vol.~70, no.~5, pp. 5131--5136, May 2021.

\bibitem{wang2021joint2}
X.~Wang, Z.~Fei, J.~Huang, and H.~Yu, ``{Joint waveform and discrete phase
  shift design for RIS-assisted integrated sensing and communication system
  under Cram{\'e}r-Rao bound constraint},'' \emph{IEEE Trans. Veh. Technol.},
  vol.~71, no.~1, pp. 1004--1009, Jan. 2021.

\bibitem{song2023intelligent}
X.~Song, J.~Xu, F.~Liu, T.~X. Han, and Y.~C. Eldar, ``{Intelligent reflecting
  surface enabled sensing: Cram{\'e}r-Rao bound optimization},'' \emph{IEEE
  Trans. Signal Process.}, vol.~71, pp. 2011--2026, May 2023.

\bibitem{chen2023ris}
Z.~Chen, P.~Chen, Z.~Guo, Y.~Zhang, and X.~Wang, ``{A RIS-based vehicle DOA
  estimation method with integrated sensing and communication system},''
  \emph{IEEE Trans. Intell. Transp. Syst.}, vol.~25, no.~6, pp. 5554--5566,
  Nov. 2024.

\bibitem{zhang2022joint}
H.~Zhang, ``{Joint waveform and phase shift design for RIS-assisted integrated
  sensing and communication based on mutual information},'' \emph{IEEE Commun.
  Lett.}, vol.~26, no.~10, pp. 2317--2321, Oct. 2022.

\bibitem{wei2022multiple}
T.~Wei, L.~Wu, K.~V. Mishra, and M.~B. Shankar, ``{Multiple IRS-assisted
  wideband dual-function radar-communication},'' in \emph{Proc. IEEE Int. Symp.
  Joint Commun. \& Sensing (JC\&S)}, Seefeld, Austria, Mar. 2022, pp. 1--5.

\bibitem{wei2022irs}
------, ``{IRS-aided wideband dual-function radar-communications with quantized
  phase-shifts},'' in \emph{Proc. IEEE Sensor Array and Multichannel Signal
  Processing Workshop (SAM)}, Trondheim, Norway, Jun. 2022, pp. 465--469.

\bibitem{luan2023robust}
M.~Luan, B.~Wang, Z.~Chang, T.~H{\"a}m{\"a}l{\"a}inen, and F.~Hu, ``{Robust
  beamforming design for RIS-aided integrated sensing and communication
  system},'' \emph{IEEE Trans. Intell. Transport. Syst.}, vol.~24, no.~6, pp.
  6227--6243, Jun. 2023.

\bibitem{liu2024ris}
R.~Liu, M.~Li, Q.~Liu, and A.~Lee~Swindlehurst, ``{SNR/CRB-constrained joint
  beamforming and reflection designs for RIS-ISAC systems},'' \emph{IEEE Trans.
  Wireless Commun.}, vol.~23, no.~7, pp. 7456--7470, Jul. 2024.

\bibitem{buzzi2022foundations}
S.~Buzzi, E.~Grossi, M.~Lops, and L.~Venturino, ``{Foundations of MIMO radar
  detection aided by reconfigurable intelligent surfaces},'' \emph{IEEE Trans.
  Signal Process.}, vol.~70, pp. 1749--1763, Mar. 2022.

\bibitem{zhang2022active}
Z.~Zhang, L.~Dai, X.~Chen, C.~Liu, F.~Yang, R.~Schober, and H.~V. Poor,
  ``{Active RIS vs. passive RIS: Which will prevail in 6G?}'' \emph{IEEE Trans.
  Commun.}, vol.~71, no.~3, pp. 1707--1725, Mar. 2023.

\bibitem{kang2024active}
Z.~Kang, C.~You, and R.~Zhang, ``{Active-IRS-aided wireless communication:
  Fundamentals, designs and open issues},'' \emph{IEEE Wireless Commun.},
  vol.~31, no.~3, pp. 368--374, Jun. 2024.

\bibitem{salem2022active}
A.~A. Salem, M.~H. Ismail, and A.~S. Ibrahim, ``{Active reconfigurable
  intelligent surface-assisted MISO integrated sensing and communication
  systems for secure operation},'' \emph{IEEE Trans. Veh. Technol.}, vol.~72,
  no.~4, pp. 4919--4931, Apr. 2023.

\bibitem{zhang2022CRAN}
Y.~Zhang, J.~Chen, C.~Zhong, H.~Peng, and W.~Lu, ``{Active IRS-assisted
  integrated sensing and communication in C-RAN},'' \emph{IEEE Wireless Commun.
  Lett.}, vol.~12, no.~3, pp. 411--415, Mar. 2023.

\bibitem{zhu2023joint}
Q.~Zhu, M.~Li, R.~Liu, and Q.~Liu, ``{Joint transceiver beamforming and
  reflecting design for active RIS-aided ISAC systems},'' \emph{IEEE Trans.
  Veh. Technol.}, vol.~72, no.~7, pp. 9636--9640, Jul. 2023.

\bibitem{yu2023active}
Z.~Yu, H.~Ren, C.~Pan, G.~Zhou, B.~Wang, M.~Dong, and J.~Wang, ``{Active RIS
  aided ISAC systems: Beamforming design and performance analysis},''
  \emph{IEEE Trans. Commun.}, vol.~72, no.~3, pp. 1578--1595, Mar. 2024.

\bibitem{zhu2024cramer}
Q.~Zhu, M.~Li, R.~Liu, and Q.~Liu, ``{Cram{\'e}r-Rao bound optimization for
  active RIS-empowered ISAC systems},'' \emph{IEEE Trans. Wireless Commun.},
  vol.~23, no.~9, pp. 11\,723--11\,736, Sep. 2024.

\bibitem{koch2008bayesian}
J.~W. Koch, ``{Bayesian approach to extended object and cluster tracking using
  random matrices},'' \emph{IEEE Trans. Aerosp. Electron. Syst.}, vol.~44,
  no.~3, pp. 1042--1059, Jul. 2008.

\bibitem{zheng2022survey}
B.~Zheng, C.~You, W.~Mei, and R.~Zhang, ``A survey on channel estimation and
  practical passive beamforming design for intelligent reflecting surface aided
  wireless communications,'' \emph{IEEE Commun. Surv. Tuts.}, vol.~24, no.~2,
  pp. 1035--1071, Feb. 2022.

\bibitem{kay1993fundamentals}
S.~M. Kay, \emph{{Fundamentals of Statistical Signal Processing: Estimation
  Theory}}.\hskip 1em plus 0.5em minus 0.4em\relax Cliffs, NJ, USA:
  Prentice-Hall., 1993.

\bibitem{grant2014cvx}
M.~Grant and S.~Boyd, ``{CVX: Matlab software for disciplined convex
  programming, version 2.1},'' 2014.

\bibitem{yang2000matrix}
X.~Yang, ``{A Matrix Trace Inequality},'' \emph{J. Math. Anal. Appl.}, vol.
  250, no.~1, p. 372, 2000.

\bibitem{yang2007mimo}
Y.~Yang and R.~S. Blum, ``{MIMO radar waveform design based on mutual
  information and minimum mean-square error estimation},'' \emph{IEEE Trans.
  Aerosp. Electron. Syst.}, vol.~43, no.~1, pp. 330--343, Jan. 2007.

\bibitem{petersen2008matrix}
K.~B. Petersen and M.~S. Pedersen, ``{The Matrix Cookbook},'' \emph{Technical
  University of Denmark}, vol.~7, no.~15, p. 510, 2008.

\bibitem{fang2023multiirsenabled}
Y.~Fang, S.~Zhang, X.~Li, X.~Yu, J.~Xu, and S.~Cui, ``{Multi-IRS-enabled
  integrated sensing and communications},'' \emph{IEEE Trans. Commun.},
  vol.~72, no.~9, pp. 5853--5867, Sep. 2024.

\bibitem{bjornson2024introduction}
E.~Bj{\"o}rnson and {\"O}.~T. Demir, \emph{Introduction to Multiple Antenna
  Communications and Reconfigurable Surfaces}.\hskip 1em plus 0.5em minus
  0.4em\relax Now Publishers, Inc., 2024.

\end{thebibliography}

\end{document}